\documentclass[twocolumn,showpacs,preprintnumbers,amsmath,amssymb,aps]{revtex4}

\usepackage{graphicx}
\usepackage{dcolumn}
\usepackage{bm}

\begin{document}

\newcommand{\be}{\begin{eqnarray}}
\newcommand{\ee}{\end{eqnarray}}

\title{Experiments of Interfacial Roughening in Hele-Shaw Flows\\with Weak
Quenched Disorder}

\author{Jordi Soriano}
\email{soriano@ecm.ub.es}

\author{Jordi Ort\'{\i}n}

\author{A. Hern\'andez-Machado}

\affiliation{Departament d'Estructura i Constituents de la Mat\`eria\\
Universitat de
Barcelona, Av. Diagonal, 647, E-08028 Barcelona, Spain}

\date{\today}

\begin{abstract}
We have studied the kinetic roughening of an oil--air interface in a forced
imbibition
experiment in a horizontal Hele--Shaw cell with quenched disorder. Different
disorder
configurations, characterized by their persistence length in the direction of
growth,
have been explored by varying the average interface velocity $v$ and the gap
spacing $b$.
Through the analysis of the {\it rms} width as a function of time, we have
measured a
growth exponent $\beta \simeq 0.5$ that is almost independent of the
experimental
parameters. The analysis of the roughness exponent $\alpha$ through the power
spectrum
have shown different behaviors at short ($\alpha_1$) and long ($\alpha_2$)
length scales,
separated by a crossover wavenumber $q_c$. The values of the measured roughness
exponents
depend on experimental parameters, but at large velocities we obtain $\alpha_1
\simeq
1.3$ independently of the disorder configuration. The dependence of the
crossover
wavenumber with the experimental parameters has also been investigated,
measuring $q_c
\sim v^{0.47}$ for the shortest persistence length, in agreement with
theoretical
predictions.
\end{abstract}


\maketitle

\section{Introduction}

The kinetic roughening of growing surfaces is a problem of fundamental interest
in
nonequilibrium statistical physics. The interest arises from theoretical,
experimental,
and numerical evidence of scale invariance and universality of the statistical
fluctuations of rough interfaces in a large variety of systems
\cite{Barabasi-Stanley,Krug-Adv-Phys-97,Meakin-1998}.  One candidate system is
the
roughening of a driven interface separating two fluids in a porous medium.  This
problem
has important practical applications, and has time and length scales easily
accessible in
the laboratory.  It allows different possible realizations, depending on the
relative
viscosities and wetting properties of the fluids involved \cite{wong-94}.

One of these possible realizations which has received considerable attention in
recent
years is {\it imbibition}, i.e. the situation in which a viscous wetting fluid
(typically
oil or water) displaces a second less--viscous, non--wetting fluid (typically
air) which
initially fills the porous medium.  The motion of imbibition interfaces can be
{\it
spontaneous}, i.e. driven solely by capillary forces, or {\it forced} externally
at
either constant applied pressure or constant injection rate.

Although there have been many experimental investigations of the scaling
properties of
imbibition interfaces in the last years, some results, particularly the
quantitative
values of scaling exponents, remain controversial.  The current situation for
the case of
spontaneous imbibition is reviewed in Ref.~\cite{Dube-00}.  The situation for
the case of
forced imbibition is summarized in Section \ref{Sec:Scaling}.  The limitations
of the
experiments, in our opinion, arise firstly from the lack of precise knowledge of
the
properties of the disorder introduced by the model porous medium, and secondly
from the
related difficulty in tuning the relative strength of stabilizing to
destabilizing forces
in the flow. In the present work we have attempted to avoid these two
limitations by
using a particular model porous medium, consisting on a Hele--Shaw cell with
precisely
designed and controlled random variations in gap spacing.

In our setup, an initially planar interface becomes statistically rough on a
mesoscopic
scale, as a result of the interplay between (i) the stabilizing effects of the
viscous
pressure field in the fluid, and the surface tension in the plane of the cell,
on long
and short length scales respectively, and (ii) the destabilizing effect of local
fluctuations in capillary pressure, arising from the random fluctuations in gap
spacing,
on short length scales. Although they have a different physical origin, the role
of local
fluctuations in capillary pressure, in our setup, is very similar to the role of
wettability defects in the {\it imperfect Hele--Shaw cell} introduced by de
Gennes
\cite{deGennes}, and studied by Paterson and coworkers \cite{Paterson}.

\begin{figure}
\includegraphics[width=8.6cm]{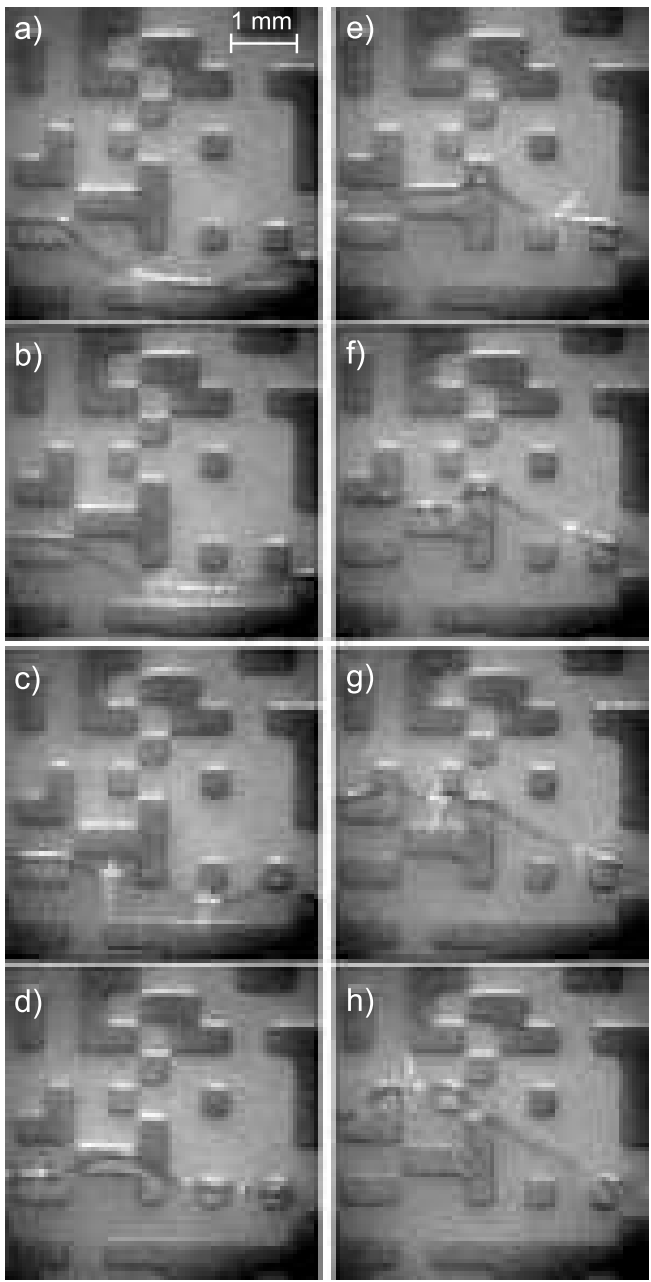}
\caption{Eight consecutive close--up views of the oil--air interface advancing
over the
disorder. In each picture the oil is driven from bottom to top. The experimental
parameters are gap spacing $b=0.36$ mm, average interface velocity $v=0.08$
mm/s, and
disorder SQ 0.40. The time interval between images is 0.8 s.}
\label{Fig:close-up}
\end{figure}

In this paper we present a systematic experimental study of forced imbibition of
our
model porous medium, by a wetting silicone oil driven at constant flow rate.
Since the
competing forces in our system are the same as in a real porous medium, the
physics
governing roughening is very similar in the two cases. The relative importance
of viscous
forces can be finely tuned by changing the injection rate of the invading fluid,
and the
strength of capillary fluctuations can be tuned by adjusting the distance
between the two
glass plates. The sequence of photographs in Fig.~\ref{Fig:close-up} provides
examples of
the resulting interfaces.

The outline of the paper is as follows. Section \ref{Sec:Scaling} reviews the
scaling
properties of rough interfaces, and their experimental characterization in
two--dimensional forced imbibition. Section \ref{Sec:Setup} describes the
experimental
setup, Section \ref{Sec:Characterization} introduces several parameters, such as
permeability and modified capillary number, useful to characterize the
experiments, and
Section \ref{Sec:DataAnalysis} explains the methodology used in data analysis.
The
experimental results are described in Section \ref{Sec:Results}, and are
analyzed and
discussed in Section \ref{Sec:Analysis}. The final conclusions are given in
Section
\ref{Sec:Conclusions}.

\section{Scaling of rough interfaces}\label{Sec:Scaling}

To be specific, let us consider a two--dimensional system with cartesian
coordinates
$(x,y)$, of lateral size $L$ in the $x$ direction and of infinite extension in
the $y$
direction.  The interface is driven in the direction of positive $y$, and its
position at
time $t$ is parametrized by the function $y=h(x,t)$. We assume that the
interface is
initially planar, $h(x,0)=0$. Let $w(l,t)$ (the {\it rms} interfacial width)
denote the
typical amplitude of transverse excursions on a scale $l$, parallel to the
interface, at
time $t$. For the complete interface:
\begin{equation}
   w(L,t) = \left[ \langle(h(x,t) -
            \widetilde{h}(t))^{2}\rangle_{x} \right]^{1/2},
\end{equation}
where $\widetilde{h}(t)=\langle h(x,t)\rangle_{x}$. The notation $\langle \ldots
\rangle_{x}$ represents an spatial average in the interval $[0, L]$, where $L$
is the
system size. The standard picture of kinetic roughening is summarized in the
dynamical
scaling assumption of Family and Vicsek \cite{Family-Vicsek-1985}:
\begin{equation}
w(L,t) = L^{\alpha}\,g \left( t/L^z \right),
\label{Eq:FV}
\end{equation}
where $\alpha$ is the {\it roughening} (or static) exponent and $z$ is the {\it
dynamic}
exponent. The scaling function $g$ is:
\begin{equation}
g(u) \sim \left\{
\begin{array}{l} u^{\alpha/z} \qquad \mbox{for} \qquad t \ll L^z, \\
\mbox{constant} \qquad \mbox{for} \qquad t \gg L^z.
\end{array}
\right.
\end{equation}
This picture assumes that the lateral correlation length of the interface
fluctuations
increases in time as $t^{1/z}$, in the $x$ direction.  For all length scales $l$
within
this correlation length the interface is rough, $w(l) \sim l^{\alpha}$.  As the
lateral
correlation length increases, the interfacial width increases correspondingly
with time
as $w \sim t^{\alpha/z}$, which defines a {\it growth} exponent $\beta =
\alpha/z$.
Finally, when the lateral correlation length exceeds the system size, $L$, at a
crossover
time $t_{\times} \sim L^z$, the interface fluctuations saturate.

One way of measuring $\alpha$ is through the analysis of the interfacial width as a
function of different system sizes $L$, i.e.  $w(L) \sim L^{\alpha}$, at saturation. From
an experimental point of view, however, it is not practical to perform experiments at
different $L$.  Usually it is prefered to measure $\alpha$ by analyzing $w$ for different
window sizes $l$, with $l \leqslant L$.  The scaling of $w(l)$ gives a {\it local}
roughness exponent $\alpha_{loc}$, which is usually identified with the {\it global}
exponent $\alpha$.  This kind of analysis must be performed with some caution, however,
because it can only provide $\alpha_{loc} \leqslant 1$ and hence fails for super--rough
interfaces ($\alpha > 1$).  Moreover, a number of problems of kinetic roughening present
intrinsic anomalous scaling, characterized by different values of local and global
exponents \cite{Lopez-96-97,Ramasco-00}.

These difficulties can be overcome by analyzing the power spectrum of the
interfacial
fluctuations, defined as
\begin{equation}\label{Power}
  S(q,t)=\langle H(q,t)H(-q,t) \rangle,
\end{equation}
where
\begin{equation}\label{Fourier}
  H(q,t)=\sum_{x}\left[ h(x,t)-\widetilde{h}(t)\right] e^{iqx}.
\end{equation}
The notation $\langle \ldots \rangle$ indicates average over disorder
configurations. The
mean width $w$ is related to $S(q,t)$ through
\begin{equation}\label{W-S}
  w^{2}(L,t)=\left( \frac{\Delta}{L} \right)^2 \sum_{q} S(q,t),
\end{equation}
where $\Delta$ is the sampling interval in the $x$ direction. For a 2--$d$
system, the
equivalent of the Family--Vicsek scaling assumption for the power spectrum reads
\cite{Lopez-96-97}:
\begin{equation}
S(q,t) = q^{-(2\alpha + 1)}\,s\left( q\,t^{1/z} \right),
\end{equation}
where the scaling function is given by
\begin{equation}
s(u) \sim \left\{
\begin{array}{l}
u^{2\alpha + 1} \qquad \mbox{for} \qquad u \ll 1, \\ \mbox{constant} \qquad
\mbox{for}
\qquad u \gg 1.
\end{array}
\right.
\end{equation}
The power spectrum can give values $\alpha > 1$ and hence is applicable to super--rough
interfaces.  On the other hand, the presence of intrinsic anomalous scaling can be
detected by a systematic shift of the power spectra computed at successive time intervals
\cite{Lopez-96-97}.

The scaling concept has allowed a classification of kinetic roughening problems
in
universality classes characterized by different families of scaling exponents.
The first
class is described by the thermal Kardar--Parisi--Zhang (KPZ) equation
\cite{KPZ-86},
which provides a local description of interfacial roughening in the presence of
an
additive white noise. The KPZ scaling exponents are $\beta = 1/3$ and $\alpha =
1/2$ in
two dimensions. If the nonlinear term of the KPZ equation is supressed, the
resulting
equation is known as the thermal Edwards--Wilkinson (EW) equation \cite{EW-82},
which
gives $\beta = 1/4$ and $\alpha = 1/2$ in two dimensions. When, instead of being
purely
thermal, the noise in the KPZ equation is supposed to depend on the interface
height $h$,
the equation displays a depinning transition. At the pinning threshold the
interface
behaviour of the complete equation with the nonlinear term (``quenched KPZ'')
can be
mapped to the directed percolation depinning (DPD) model, whose scaling
exponents in two
dimensions are $\beta = \alpha \simeq 0.633$.  Suppressing the nonlinearity
yields the
``quenched EW'' equation, for which $\beta \simeq 0.88 $ and $\alpha \simeq 1$
in two
dimensions \cite{Leschhorn-96-97}.

In the case of imbibition, there are two important issues which make the above
classification of limited applicability.  The first one is the quenched nature of the
disorder. It has been argued that for very large driving the disorder may still be
considered as a fluctuating in time (thermal) noise, but in general the disorder must be
treated as a static, quenched noise. The second issue is the nonlocal character of the
dynamics, due to fluid transport in the cell \cite{krug-91,he-92}.  This issue has
received important attention recently, with the introduction of imbibition models which
take fluid transport explicitly into account, giving rise to nonlocal interfacial
equations \cite{ganesan-98,Dube-teoric-99,Dube-teoric-00,Aurora-EPL-01}.

The models presented in \cite{ganesan-98,Dube-teoric-99,Dube-teoric-00,Aurora-EPL-01} are
consistent with the well known macroscopic equations of the problem (Darcy's law and
interfacial boundary conditions).  They differ in the way the noise is included in the
equations and in the noise properties.  The starting point of Ganesan and Brenner
\cite{ganesan-98} is a random field Ising model.  The permeability is taken spatially
uniform and the noise exhibits long range spatial correlations.  In the case of forced
imbibition, and based on a Flory--type scaling, the model predicts that the roughness
exponent $\alpha$ depends on the capillary number Ca, with asymptotic values $\alpha =
3/4$ for the smallest drivings and $\alpha = 1/2$ for the largest drivings. The
approaches of Dub{\'e} {\it et al.} \cite{Dube-teoric-99,Dube-teoric-00} and
Hern\'andez--Machado {\it et al.} \cite{Aurora-EPL-01} are both based in a conserved
Ginzburg-Landau model, where the noise is introduced in a fluctuating chemical potential
or in the mobility, respectively, without long range correlations. Dub{\'e} {\it et al.}
\cite{Dube-teoric-99,Dube-teoric-00}, however, consider only the case of spontaneous
imbibition, which could give results very different from the case of forced imbibition,
specially for the growth exponent $\beta$.  By numerical integration they obtain
$\alpha=1.25$ and $\beta=0.3$.  Hern\'andez--Machado {\it et al.} \cite{Aurora-EPL-01}
study the case of forced imbibition, and predict a different scaling of the short and
long length scales, with exponents $\beta_1 = 5/6, \alpha_1 = 5/2$ in the former regime,
and $\beta_2 = 1/2, \alpha_2 = 1/2$ in the latter.

A common feature of these models, pointed out in Refs.
\cite{Dube-teoric-99,Dube-teoric-00,Aurora-EPL-01}, is the presence of a new lateral
length scale, $\xi_c$, related to the interplay of interfacial tension and liquid
conservation.  For $q \gg 1/\xi_c$ the dominant stabilizing contribution is the
interfacial tension in the plane of the cell, while for $q \ll 1/\xi_c$ it is the fluid
flow.  The interplay leads to a dependence of $q_c$ ($=1/\xi_c$) on $v$ of the form $q_c
\sim v^{1/2}$.

As mentioned in the introduction, the experimental characterization of the scaling
properties of interfaces in two--dimensional forced imbibition remains conflicting. Most
experiments have been conducted in model porous media consisting of air--filled
Hele--Shaw cells packed with glass beads.  In a first experiment of this sort on
water--air interfaces, carried out by Rubio {\it et al.} \cite{Rubio-89}, a roughening
exponent $\alpha = 0.73 \pm 0.03$ was measured, independent of Ca and bead size, for Ca
values in the range $10^{-3}$ -- $10^{-2}$.  A controversial reanalysis of their data by
Horv\'ath {\it et al.} gave $\alpha = 0.91 \pm 0.08$ \cite{Horvath-90,Rubio-90}, which
these authors compared to $\alpha = 0.88 \pm 0.08$ obtained in their own replication of
the experiment. In a subsequent work \cite{Horvath-91}, using glycerol instead of water,
Horv\'ath {\it et al.} reported a clear power law growth regime with a growth exponent
$\beta \simeq 0.65$, and different values $\alpha \simeq 0.81$ and $\alpha \simeq 0.49$
at short and long wavenumbers, respectively, at saturation.  The last set of experiments
of this kind is due to He {\it et al.} \cite{he-92}, who explored a very large range of
Ca (from $10^{-5}$ to $10^{-2}$) and found large fluctuations of the roughness exponent
in the saturation regime, between $0.65$ and $0.91$. In these experiments $w$ was shown
to fluctuate wildly during growth, and it was impossible to give a value of the growth
exponent $\beta$.

Other experiments have been directed to characterize the statistical properties
of the
avalanches displayed by imbibition fronts at sufficiently small Ca. The results
are also
conflicting.  The first study \cite{Horvath-avalanches-91} was done on the same
air-glycerol interfaces of Ref.~\cite{Horvath-91}, and gave a power law
distribution of
avalanche sizes.  The second one \cite{Dougherty-Carle-98}, on air--water
interfaces,
reported an exponential distribution.

In our experiments, the possibility of tuning the different competing forces has
allowed
an accurate measurement of the growth exponent $\beta$, by enlarging the growth
regime
before saturation.  Fine tuning of the forces has also allowed to measure the
crossover
length $\xi_c$, and its dependence on velocity $v$.

\section{Experimental Setup} \label{Sec:Setup}

\begin{figure}
\includegraphics[width=8.6cm]{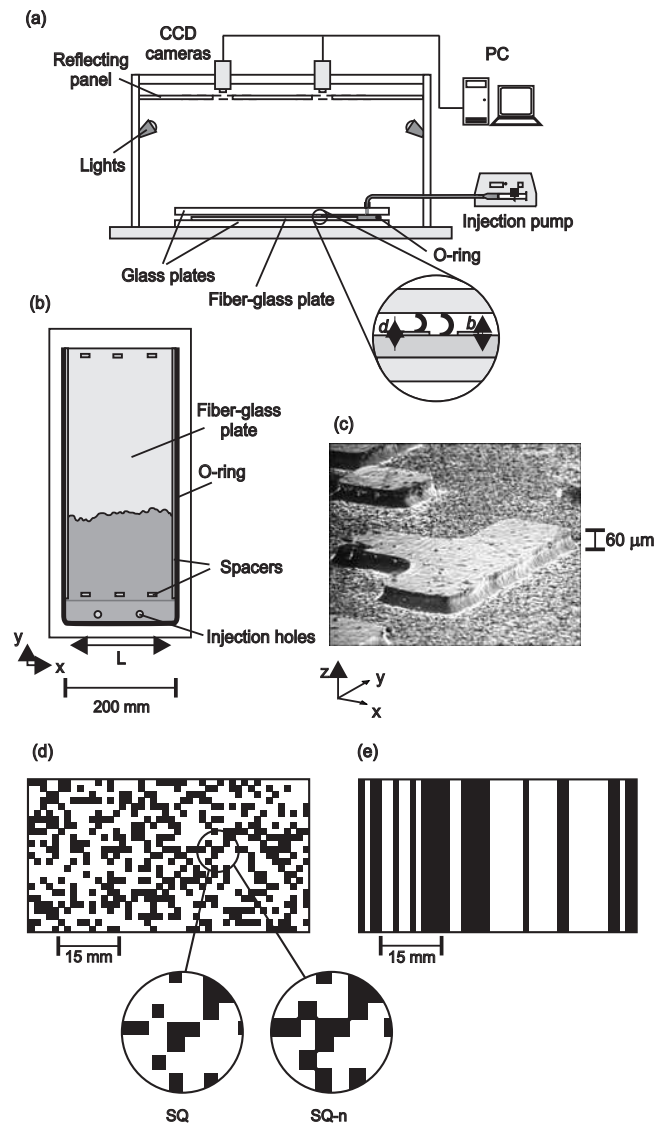}
\caption{Sketch of the experimental setup. a) Side view. b) Top
view. c) SEM image of the copper islands on the fiber--glass plate. d) and e) Partial
views of the disorder pattern for d) squares (SQ, SQ-n) and e) tracks (T). The copper
islands are the black regions. The blow--ups in d) show the detailed structure of the
disorder pattern for SQ and SQ-n.} \label{Fig:set-up}
\end{figure}

In our experiments a silicone oil (Rhodorsil 47 V) displaces air in a horizontal
Hele-Shaw cell, $190 \times 550$ mm$^{2}$, made of two glass plates 20 mm thick.
The oil
has kinematic viscosity $\nu=50$ mm$^2$/s, density $\rho=998$ kg/m$^{3}$, and
surface
tension oil--air $\sigma=20.7$ mN/m at room temperature. Fluctuations in the gap
thickness are provided by a fiber-glass substrate, fixed on the bottom glass
plate,
containing a large number of copper islands which randomly occupy the sites of a
square
grid. The height of the islands is $d=0.06 \pm 0.01$ mm. The gap spacing $b$,
defined as
the separation between the substrate and the top plate, is set by placing
several
calibrated spacers on the perimeter of the substrate, over the disorder, as
shown in
Fig.\ \ref{Fig:set-up}. We have used gap thicknesses in the range $b=0.16-0.75$
($\pm
0.05$) mm.

The disorder pattern is designed by computer and manufactured using printed
circuit
technology. In order to clearly identify the oil--air interface when it moves
over the
copper islands, the plates have been chemically treated to accelerate copper
oxidation.
Otherwise the copper islands are too bright to recognize the interface contour.
The
silicone oil wets perfectly both fiber--glass and copper, and no differences in
the
wetting properties due to the oxidation protocol have been observed.

For each experiment, the glass plates are cleaned with soap and water and rinsed
with
distilled water and acetone. The fiber--glass plate is cleaned using
blotting--paper
only, which leaves a thin layer of oil to avoid further oxidation of the copper
surface,
ensures a complete wetting in both fiber--glass and copper, and improves the
interface
contrast in the captured images. Next, the fiber--glass plate is fixed over the
bottom
glass plate using a thin layer of removable glue. The cell is completed by
placing the
spacers, the upper glass plate, and the o--ring around the fiber--glass plate
(Fig.\
\ref{Fig:set-up}). Finally, the two glass plates are firmly clamped together, to
ensure
the homogeneity of the gap spacing and the tightness of cell sides closed by the
o--ring.

The oil is injected at constant flow-rate using a syringe pump Perfusor ED-2.
The syringe
pump can be programmed to give flow--rates $Q$ in the range 1--299 ml/h with
less than 2
\% fluctuations around the nominal value. The oil enters the cell through two
wide holes,
drilled on the top plate near one end of the cell. The other end of the cell is
left
open. To start the experiment with as flat an interface as possible, the oil is
first
slowly injected on a transverse copper track, on the fiber-glass plate, which is
2 mm
ahead the disorder pattern. Next, the syringe pump is set to its maximum
injection rate
until the whole interface has reached the disorder (about 3 s later). The pump
is then
set to the nominal injection rate of the experiment, and $t=0$ is defined as the
time at
which the average height of the interface (measured on the images) reaches the
preset
nominal velocity.

The oil--air interface evolution is monitored using two JAI CV-M10BX progressive
scan CCD
cameras (each camera acquiring half side of the cell). The 1/2" CCD sensor
contains 782
(H) $\times$ 582 (V) pixels. In our experiment we have used an exposure time of
1/25 s
in order to minimize the illumination. Each camera is equipped with a motorized
zoom lens
Computar M10Z1118MP with a focal length in the range 11--110 mm (1:10 zoom
ratio). The
cameras are connected to an Imaging Technology PCVision frame grabber installed
in a
personal computer. A Visual Basic application controls both cameras and stores
the images
for further analysis.

\begin{figure}
\includegraphics[width=8.6cm]{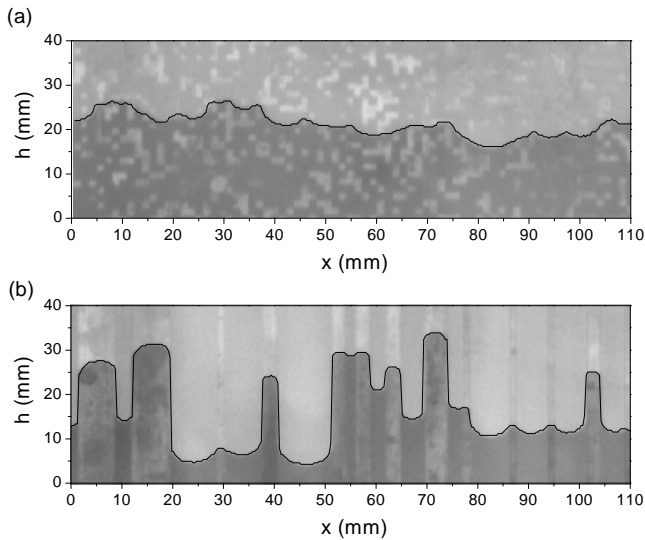}
\caption{Interface profiles for two types of disorder
configurations, SQ 1.50 (a) and T 1.50 (b), compared with the original digital
images.
The copper islands are the brighter spots of the image background. The
experimental
parameters are $b=0.36$ mm and $v=0.08$ mm/s in the two cases.}
\label{Fig:image}
\end{figure}

The images are taken with a spatial resolution of 0.37 mm per pixel, a size of
$768
\times 574$ pixels, and 256 grey scale levels per pixel. The acquisition is
logarithmic
in time, with temporal increments that vary from 0.33 s to 180 s. Between 100
and 300
images are taken per experiment. Because the background is the same for all the
images
captured with the same camera, the interface is enhanced by subtracting the
first image
to all other images, and thresholding the result to get a black and white
contour of the
interface. The contour is resolved with 1--pixel accuracy using edge detection
methods,
and data are stored for further analysis. This method is automatically
performed, with an
error in the interface recognition comparable to the width of the oil--air
meniscus,
about one half the gap width. Fig.\ \ref{Fig:image} presents two examples of the
digitized interfaces, compared with the original images.

\section{Characterization of the experiments}\label{Sec:Characterization}

\subsection{Disorder properties}\label{Subsec:disorder}

\begin{figure}
\includegraphics[width=8.6cm]{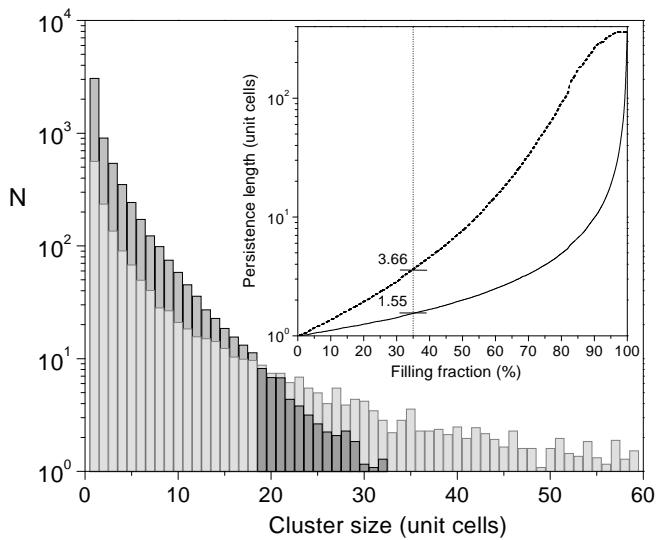}
\caption{Statistical distribution
of the copper clusters for a filling fraction $f=0.35$. Clusters with only nearest
neighbour contacts (SQ) are represented in dark grey. Clusters with nearest and
next--nearest neighbour contacts (SQ-n) are represented in light grey. The inset shows
the persistence length of the disorder $\widetilde{l}$ as a function of the filling
fraction $f$. The solid and dashed lines correspond to the SQ and SQ-n cases
respectively. The vertical line indicates the filling fraction used in our experiments.}
\label{Fig:noise}
\end{figure}

Three kinds of disorder patterns have been used. Two of them are obtained by random
selection of the sites of a square lattice (Fig.\ \ref{Fig:set-up}(d)). In the first of
them (SQ) we allow nearest neighbour connections only, leaving next--nearest neighbours
separated by 0.08 mm. In the second one (SQ-n) both nearest neighbour and next--nearest
neighbour connections are allowed. The third kind of disorder pattern (T) is formed by
parallel tracks, continuous in the $y$ direction and randomly distributed along $x$
(Fig.\ \ref{Fig:set-up}(e)). The filling fraction $f$ (fraction of lattice sites occupied
by copper) is 35 \% in the three cases. However, $\widetilde{l}$, the persistence length
of the disorder in the direction of growth, is very different in the three different
disorder patterns. This length is measured in the following way: (i) We consider every
site $x$ of the lattice along the lateral direction, and measure the average length
$l(x)$ of the island formed by the connected copper sites at $x-\Delta x$, $x$, and
$x+\Delta x$, where $\Delta x$ is the lattice spacing. (ii) We average $l(x)$ over the
lateral direction in the interval $(0,L)$. As can be seen in the inset of Fig.\
\ref{Fig:noise}, $\widetilde{l}$ increases by a factor $2.4$ when changing from SQ to
SQ-n, and up to the total length of the cell for T. Another characteristic of the
disorder is the cluster size, which gives the number of disorder unit cells of the copper
aggregations. The statistical distribution of copper clusters is shown in the main plot
of Fig.\ \ref{Fig:noise}.

We have used two values of the lattice spacing in the lateral direction, $\Delta
x = 1.50 \pm 0.04$ mm and $0.40 \pm 0.04$ mm. From here on we refer to the
disorder used in a given experiment by SQ, SQ-n, or T,
followed by the lateral size in mm of the disorder unit cell, $1.50$ or $0.40$.

\subsection{Permeability}

To characterize the fluid flow through our Hele--Shaw cell with disorder, we
have
determined the permeability of the cell for the different disorder patterns as a
function
of the gap spacing. The experimental set up used for this purpose is similar to
the one
shown in Fig.\ \ref{Fig:set-up}, but the injection system has been replaced by a
constant
pressure device. It consists of an oil column of adjustable constant height in
the range
from 200 $\pm$ 2 mm to 1000 $\pm$ 5 mm. The permeability is determined by
measuring the
oil--air interface average velocity for different applied pressures (heights of
the oil
column) and using Darcy's law,
\begin{equation}\label{Darcy}
  \vec{v} = - \frac{k}{\mu}{\vec{\nabla}}p,
\end{equation}
where $\vec{v}$ is the interface velocity, $k$ the permeability, $\mu$ the
dynamic
viscosity, and ${\vec{\nabla}}p$ the pressure gradient.

\begin{figure}
\includegraphics[width=8.6cm]{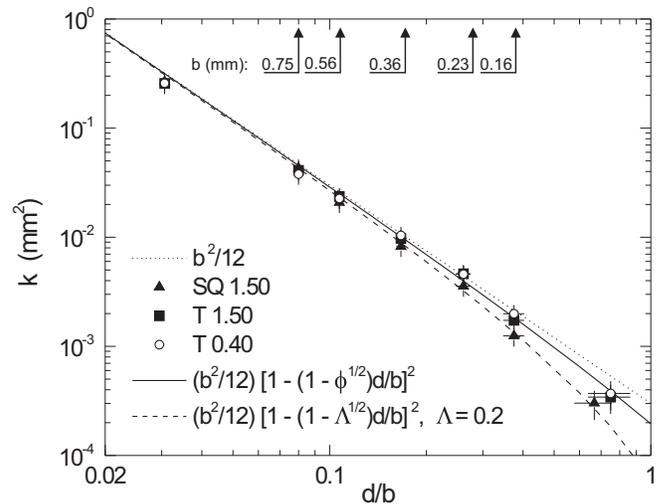}
\caption{Measurements of the permeability, $k$, as a function of
the disorder strength, $d/b$, for $d=0.06$ mm. The dotted line is the
permeability of an
standard Hele--Shaw cell (without disorder), and the dashed and solid lines are
fits to
the experimental data for SQ and T disorder configurations, respectively. The
arrows
point to the values of $d/b$ for the gap spacings used in the experiments.}
\label{Fig:permeability}
\end{figure}

Fig.\ \ref{Fig:permeability} shows the results obtained. At large gap spacings,
$d/b
\rightarrow 0$, the disorder has no effect on the fluid flow, and the
permeability tends
to the expected value for an ordinary Hele--Shaw cell, $k_{0}=b^{2}/12$,
independently of
the disorder configuration. At very small gaps, $d/b \rightarrow 1$, the
permeability decreases, and tends to a non--zero value $k_{1}$ for $d/b=1$, that
clearly depends on the
disorder configuration. We have found it convenient to write $k_{1}$ in the form
\begin{equation}\label{k1}
  k_1=d^2{\Lambda}/12,
\end{equation}
where $\Lambda$ is a function that depends on the porosity $\phi$ and the
geometry of the disorder. The simplest functional form that interpolates between
these two limits can be written as:
\begin{equation}\label{Perm-general}
  k=\frac{b^{2}}{12}\left(1-(1-{\Lambda}^{1/2})\frac{d}{b}\right)^{2}.
\end{equation}
The coefficient $\Lambda$ can be obtained in general from a
fit of the permeability to the experimental data.

In the particular case of disorder T and $d/b=1$, an analytic expression of
$\Lambda$ can be derived by recognizing that the cell in this case is formed by
a parallel array of rectangular capillaries.  Following Avellaneda {\it et al.}
\cite{Avellaneda91}, in this geometry $\Lambda$ is directly the porosity $\phi$,
which in
the limit $d/b=1$ is given simply by $\phi=1-f$.  Hence, since we have $f=0.35$,
we
obtain $\Lambda=0.65$. This result fits well the experimental data not only in
the limit
$d/b=1$ but also in the whole range of $d/b$ (solid line in Fig.\
\ref{Fig:permeability}). Notice also that the results presented in Fig.\
\ref{Fig:permeability} show that there are no important differences between T
1.50 mm and
T 0.40 mm. This observation generalizes the theoretical result that in the limit
$d/b=1$
the width of the rectangular capillaries does not modify the permeability
\cite{Avellaneda91}. For SQ and SQ--n disorder, due to the difficulty of finding
a
general expression for $k_{1}$ (see Refs.~\cite{Koponen96,Koponen97}), we have
fitted $k$
to our experimental results (dashed line in Fig.\ \ref{Fig:permeability}) and
have
obtained a numerical value $\Lambda=0.20$.

\subsection{Capillary pressure}

Assuming local thermodynamic equilibrium, the capillary pressure jump at the
interface is
given by
\begin{equation}\label{P}
  p^{+}-p^{-}=\sigma(\kappa_{\parallel}+\kappa_{\perp}),
\end{equation}
where $\kappa_{\parallel}$ and $\kappa_{\perp}$ are the main curvatures of the
interface,
in the plane of the cell and perpendicular to it, respectively.

$\kappa_{\parallel}$ varies from $0$ for $d/b \ll 1$, to $2/r$ for $d/b \simeq
1$, where
$r \simeq 1.75$ mm is the average diameter of the copper obstacles. In the range
of gap
spacings explored we have measured curvatures in the range $3.3 \times 10^{-3}
\leqslant
\kappa_{\parallel} \leqslant 1.15$ mm$^{-1}$.  Notice that $\kappa_{\parallel}
\ll
\kappa_{\perp}$ in all the range of gap spacings used in the experiments.

In ordinary Hele--Shaw flows (without disorder) $\kappa_{\perp}$ is roughly the
same in
all points of the interface and therefore adds only a constant contribution to
the
pressure jump. In our case, however, when the interface is over the copper
islands (gap
thickness $b-d$) we have:
\begin{equation}\label{P-CU}
  \kappa_{\perp}=\frac{2}{b}\left(\frac{1}{1-d/b}\right),
\end{equation}
and when it is over the fiber--glass substrate (gap thickness $b$):
\begin{equation}\label{P-FG}
  \kappa_{\perp}=\frac{2}{b},
\end{equation}
assuming complete wetting.

This difference in curvature, given by
\begin{equation}\label{Psi}
 \Psi=\frac{2}{b}\left(\frac{d/b}{1-d/b}\right),
\end{equation}
makes the interface experience a capillary instability when it passes from one
gap
spacing to the other. In the range of gap spacings studied, $0.16 \leqslant b
\leqslant
0.75$ mm, we get  $7.5 \geqslant \Psi \geqslant 0.23$ mm$^{-1}$. We have
verified that
for gap spacings $b \gtrsim 2$ mm, which correspond to $\Psi \lesssim 0.031$
mm$^{-1}$,
the fluctuations in capillary pressure are no more sufficient to roughen the
interface
appreciably.

\subsection{Modified capillary number}\label{subsec:Ca}

Once the permeability of the cell has been characterized, we can
introduce a dimensionless number to describe the relative strength of
viscous to capillary forces. The simplest number that relates viscous
and capillary forces is the {\it capillary number} Ca$=\mu v /
\sigma$. In order to account for the properties of the disorder,
which are not contained in the previous definition of Ca, it is
customary to introduce a {\it modified capillary number}. For an
ordinary Hele-Shaw cell (without disorder), the modified capillary
number, which we call Ca$^*$, comes out from the dimensionless
form of the Hele--Shaw equations \cite{Homsy87}:
\begin{equation}\label{CaMod-HeleShaw}
   \mbox{Ca}^* = \mbox{Ca} \cdot  12 \left(\frac{L}{b}\right)^2.
\end{equation}
To define a modified capillary number Ca' for our
particular cell with disorder, we consider on one side
the average viscous pressure drop across
the cell, given by
\begin{equation}\label{viscous-drop}
 \delta p_{vis}=L \cdot |\nabla p_{vis}|,
\end{equation}
with $\nabla p_{vis}=-v \mu / k$ (Darcy's law). $L$, the cell width,
provides the macroscopic length scale. On the other side, a measure of the
capillary pressure drop is given by
\begin{equation}\label{capillary-drop}
 \delta p_{cap}= \sigma \left(\frac{2}{b-d}-\frac{2}{b} + \frac{1}{L}
 \right).
\end{equation}
The last contribution accounts for the curvature of the interface in the plane of the
cell, and is relevant only when $d/b \to 0$, i.e. when the destabilizing role of the
disorder vanishes.

Defining Ca' $= \delta p_{vis} / \delta p_{cap}$ and using (\ref{Perm-general}),
(\ref{viscous-drop}) and (\ref{capillary-drop}), we get
\begin{equation}\label{CaMod}
   \mbox{Ca'}= \mbox{Ca}
  \cdot \frac{12L}{\displaystyle
  b^{2}\left(1-(1-{\Lambda}^{1/2})\frac{d}{b}\right)^{2}
  \left(\frac{2}{b-d}-\frac{2}{b}+\frac{1}{L}\right)}.
\end{equation}

\begin{figure}
\includegraphics[width=8.6cm]{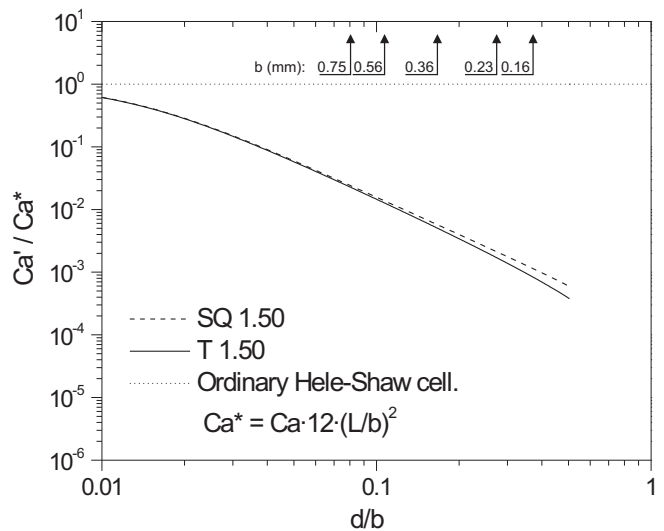}
\caption{Ratio of the modified capillary number for the cell with disorder, Ca',
to the modified
capillary number for an ordinary Hele-Shaw cell, Ca$^*$, as a function
of the disorder strength, $d/b$. The dashed and solid lines
correspond to disorders SQ 1.50 and T 1.50. The dotted line
represents the asymptotic limit Ca'$=$Ca$^*$.  The arrows point to the
values of $d/b$ for the gap spacings used in the experiments.}
\label{Fig:ca}
\end{figure}

The ratio Ca'/Ca$^*$ as a function of $d/b$ is shown in Fig.\ \ref{Fig:ca} for disorders
SQ and T. When $d/b \to 0$ the destabilizing role of the disorder is negligible, as
expected, and Ca'/Ca$^*$ tends to $1$ (ordinary Hele-Shaw cell). As $d/b$ increases, the
increasing strength of the disorder is manifest in the progressive decrease of
Ca'/Ca$^*$. It is important to notice that our definition of Ca' is not valid for
$d/b=1$, because the flow would be essentially different when the free gap disappears. In
the range of $d/b$ explored in our experiments the permeability of the cell remains
always close to that of an ordinary Hele--Shaw cell (Fig.\ \ref{Fig:permeability}), and
the decrease of Ca'/Ca$^*$ is essentially due to the capillary forces associated with the
menisci in the $z$ direction.

In the range of gap spacings experimentally explored, $0.16 \leqslant b \leqslant 0.75$
mm, the ratio between the free gap $b-d$ and the total gap $b$ varies between 62\% and
92\%. These large ratios, combined with the high viscosity and small surface tension of
the silicone oil compared with other fluids (i.e. water), are responsible for the large
values of Ca' in our experiments, in comparison to the values reported for pure porous
media. Considering for example the disorder T 1.50 mm, Ca' varies from a value $1.33$
(for $b=0.16$ mm and the minimum interface velocity, $0.04$ mm/s) to a value $17.0$ (for
$b=0.75$ mm and the maximum interface velocity, $0.4$ mm/s).

\section{Data Analysis}\label{Sec:DataAnalysis}

The wetting of the lateral gap spacers by the invading oil changes the physics at the two
ends of the interface.  To minimize this disturbance, which is particularly important at
large gap spacings, we have disregarded 8 mm at each side of the cell, thereby reducing
the measured interface from 190 to 174 mm in the $x$ direction. The final number of
pixels of the interface after this correction is reduced from 515 to 470.  For data
analysis convenience these pixels are converted into $N=512$ equispaced points, through a
linear interpolation. The final spacing between two consecutive points is $\Delta = L/N =
0.34$ mm . Although the linear interpolation introduces an artificial resolution 7 \%
larger than the resolution of the original image, the increase does not affect the final
analysis.

The interfaces measured at the smallest gap spacings and velocities may present
overhangs. These multivaluations, rather exceptional, have been eliminated by taking for
each value of $x$ the corresponding largest value of $h(x)$.

In addition, we have forced periodic boundary conditions for $h(x,t)$ by subtracting the
straight line connecting the two ends of the interface. This procedure is well documented
in the literature of kinetic roughening \cite{Simonsen98}. The linear correction imposed
to the interfaces has significant effects on the power spectrum. The Fourier spectrum of
an interface which is discontinuous at the two end points is dominated by an overall
behaviour of the form $q^{-2}$.  Forcing periodic boundary conditions eliminates the
overall slope -2 \cite{Schmittbuhl95}. Moreover, the analysis of $w(l)$ is insensitive to
the linear correction of the interface (except for values of $l$ comparable to the system
size) and gives results consistent with those obtained from $S(q,t)$ only for interfaces
with periodic boundary conditions.

Since the maximum width of the meniscus is only one half the gap width (in
conditions of
complete wetting) the interface can be considered one--dimensional at the length
scale of
the copper islands. The resolution of the interface fluctuations in the $y$
direction is
$\pm$1 pixel at a given point. Nevertheless, the measurement of the global
interfacial
width $w$ is much more precise because the width is an average over the $N=512$
points of
the interface.

Given that, according to our choice of the time origin, the whole interface is already
inside the disorder at $t=0$, $w(0) \neq 0$. For this reason, we have decided to
characterize the interface fluctuations by the {\it subtracted width} $W$, defined by
$W^{2}(t)=w^{2}(t) - w^{2}(0)$ \cite{Barabasi-Stanley,Tripathy00}. Notice that, since the
data analysis is based on power law dependencies, short times are very sensitive to the
definition of $t=0$ and to the value $w(0)$. After analyzing the data for different
definitions of $t=0$, and checking the influence of subtracting $w(0)$, we have found
that the analysis based on the subtracted width is the most objective and less sensitive
to the details of the experimental procedure. The error bars shown in the $W(t)$ plots
indicate the dispersion of the different individual experiments with respect to the
average curve plus the uncertainty in the determination of $t=0$.

Finally, the crossover time $t_{\times}$ has been measured on the $W(t)$ log-log
plots as
the time when the power law with slope $\beta$ crosses the horizontal straight
line that
corresponds to the average value of the interfacial width at saturation,
$W_{s}$.

\section{Experimental results}\label{Sec:Results}

The parameters explored in our experiments are summarized in Table\
\ref{TAB:exp-list}.
The minimum velocity selected in the experiments, $v=0.04$ mm/s (which will be
taken as
reference unit for interfacial velocities), has been chosen to ensure that the
interface
is always single--valued for gap spacings $b \gtrsim 0.36$ mm. We have used
three
different disorder realizations for SQ and SQ-n and four for T. For each
disorder
realization and each set of experimental parameters ($v$, $b$) we have carried
out three
runs for SQ and SQ-n, and two runs for T.

\begin{table*}
\begin{center}
\begin{tabular}{|c|c|c|}
\hline Disorder & Gap spacing (mm) &  Interface velocity ($V=0.04$ mm/s)
\\ \hline \hline SQ 1.50 & $0.16, 0.23, 0.36, 0.56, 0.75$ &
$2V$ \\ SQ 1.50 & $0.36$ & $V, 2V, 5V, 7V, 10\,V$ \\ \hline SQ-n 1.50 & $0.36$ &
$V,2V,10\,V$ \\ \hline T 1.50 & $0.16, 0.36, 0.56, 0.75 $ & $2V$ \\ T 1.50 &
$0.36$ & $V,
2V, 5V, 10\,V$ \\ \hline SQ 0.4 & $0.36$ & $2V$ \\ \hline T 0.4 & $0.36$ & $2V$
\\ \hline
\end{tabular}
\caption{Summary of the parameters explored for each kind of disorder.}
\label{TAB:exp-list}
\end{center}
\end{table*}

\subsection{Variable velocity}

\subsubsection{SQ 1.50 mm}\label{Subsec:SQ}

\begin{figure}
\includegraphics[width=8.6cm]{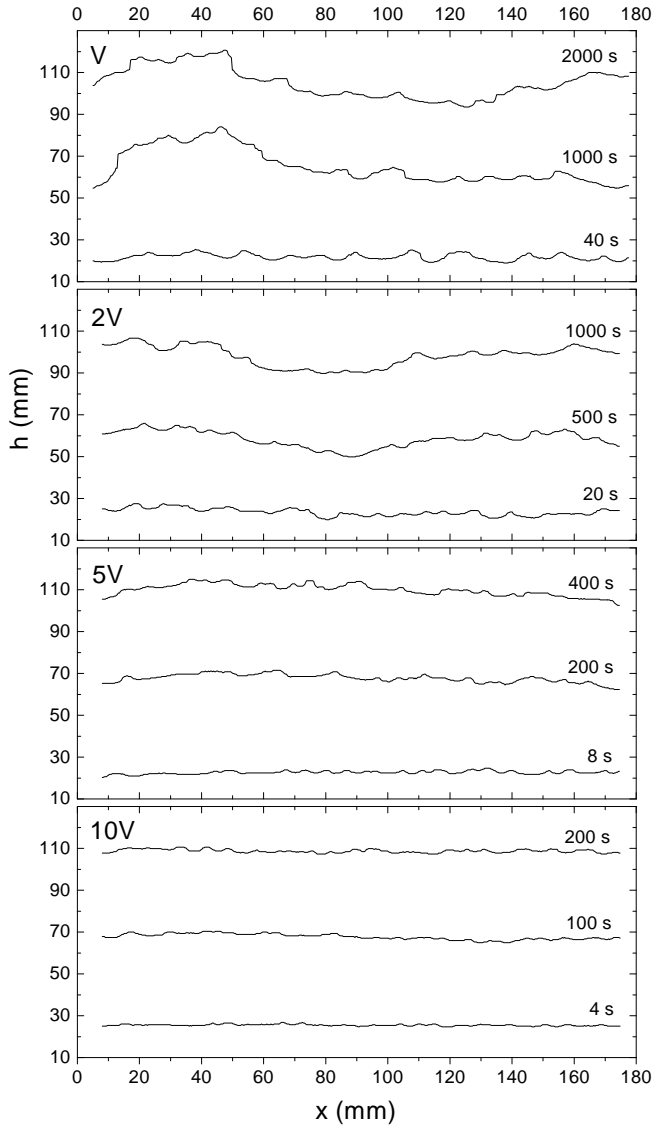}
\caption{Sequence of interfaces at different velocities, for $b=0.36$ mm and
disorder SQ
1.50, corresponding to Ca'$= 1.85$, $3.69$, $9.24$, and $18.5$. The disorder
realization
is the same in the four sequences. The reference velocity is $V=0.04$ mm/s.}
\label{Fig:int-sq-v}
\end{figure}

We have started the study exploring five different velocities, from $V$ to
$10\,V$, using
an intermediate gap spacing $b=0.36$ mm. We have not selected velocities lower
than $V$
to avoid overhangs and trapped air. An example of the interfaces obtained at
different
velocities is presented in Fig.\ \ref{Fig:int-sq-v}. All the experiments shown
correspond
to the same disorder configuration, and the average position of the selected
interfaces
is also the same for the different velocities. The most noticeable aspect of the
sequence
of interfaces is that the short length scales are not affected significantly by
the
velocity, as opposite to the long length scales, which become progressively
smoother as
the velocity increases.

\begin{figure}
\includegraphics[width=8.6cm]{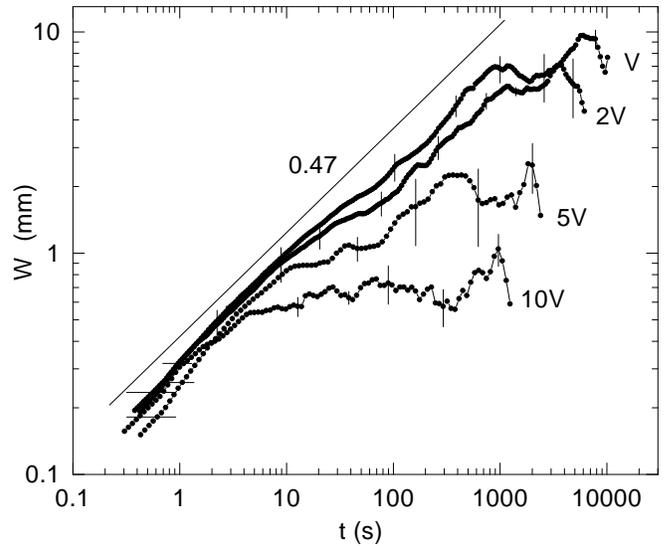}
\caption{Interfacial width, $W(t)$, for disorder SQ 1.50, gap
spacing $b=0.36$ mm, and four different velocities. The straight line drawn to
guide the
eye is the power law $W \sim t^{0.47}$.} \label{Fig:w-t-sq-v}
\end{figure}

Fig.\ \ref{Fig:w-t-sq-v} shows the $W(t)$ plot for four different velocities. We
have
omitted the curve for $7V$ because it is too close to the $10\,V$ curve. All
curves
display a power law growth regime with approximately the same growth exponent
$\beta=0.47
\pm 0.04$, independent of the velocity. Saturation times and saturation widths
depend
clearly on the velocity, both decreasing at increasing velocities. This general
behaviour
is not so clear for the two smallest velocities, for which the corresponding
$t_{\times}$
and $W_{s}$ are practically the same. This can be a consequence of the fact that
the
interface at very low velocity gets locally pinned at the end of the copper
obstacles,
hindering an arbitrary deformation of the interface. The strong fluctuations of
$W$ that
can be observed both during growth and at saturation are also remarkable. These
fluctuations have two origins: the intrinsic metastabilities in systems with
quenched--in
disorder \cite{he-92}, and to a lesser degree the inhomogeneities in the gap
thickness
caused by small deformations in the fiber-glass substrate or small differences
in the
height of the copper obstacles. To smooth out the fluctuations it would have
been
necessary to increase the number of disorder configurations and runs to a number
that
would be prohibitive. Nevertheless, the largest fluctuations usually appear deep
in
saturation and do not change the results presented here.

\begin{figure}
\includegraphics[width=7.2cm]{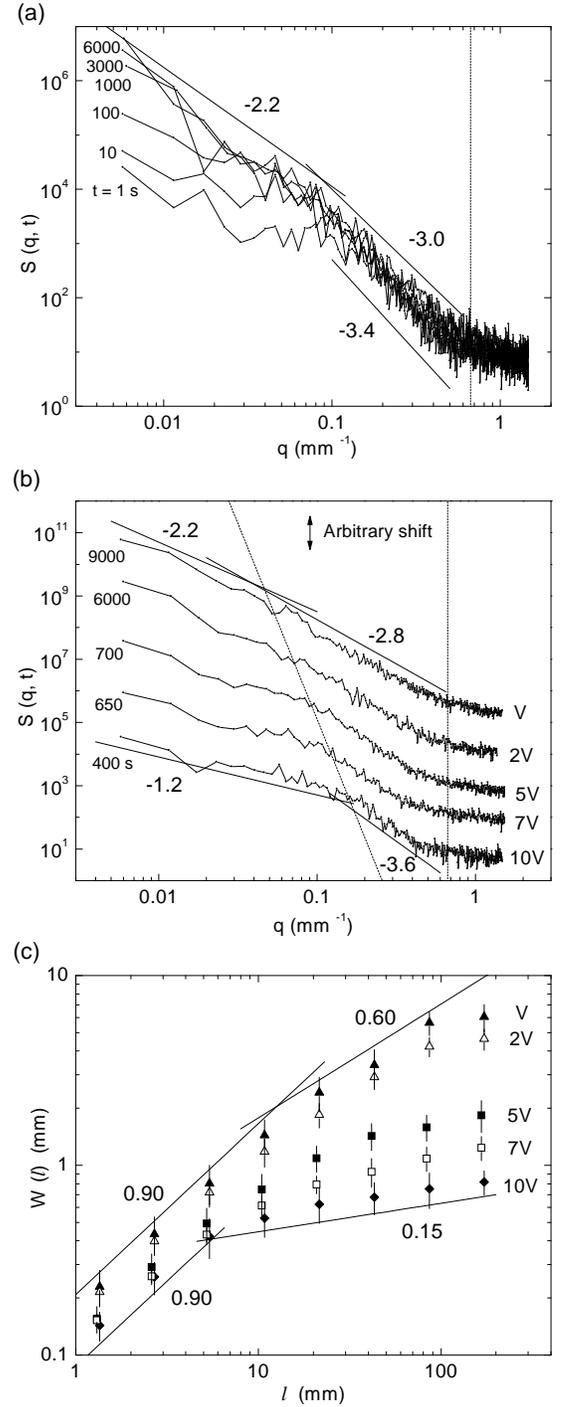}
\caption{Determination of the roughness exponents for experiments at five
different
velocities, with $b=0.36$ mm and disorder SQ 1.50. a) Example of the temporal
evolution
of the power spectrum for velocity $2V$ (Ca'=$3.69$). b) Power spectra at
saturation for
the five velocities studied. The curves have been shifted in the vertical
direction for
clearness. c) Analysis of the local width, $W(l)$. The vertical lines in (a) and
(b) give
the value of $q$ associated with the spatial scale of the disorder. The tilted
line
crossing the power spectra in (b) fits the values of the crossover wavenumber
$q_c$ at
the different velocities. The other straight lines are power law fits from which
the
roughness exponents can be deduced.} \label{Fig:pow-sq-v}
\end{figure}

Fig.\ \ref{Fig:pow-sq-v} shows the analysis of the interface fluctuations
through the
power spectrum. An example of the temporal evolution of the spectrum, which
corresponds
to the experiments at velocity $2V$, is presented in Fig.\
\ref{Fig:pow-sq-v}(a). Since
larger length scales become progressively saturated with time, the spectrum at
short
times displays a power law decay for large $q$ only. As time increases the power
law
extends to smaller $q$ until a second power law behaviour emerges, which reaches
the
smallest $q$, corresponding to the system size $L$, at saturation. The first
power law
regime (large $q$) will be characterized with a roughness exponent $\alpha_{1}$,
and the
second one (small $q$) with a roughness exponent $\alpha_{2}$. In the range of
parameters
explored we have always observed $\alpha_{1} > \alpha_{2}$. It is interesting to
notice
that the first regime shows a slightly time--dependent behaviour. $\alpha_{1}$
is
progressively smaller as time increases, and reaches a constant value at
saturation. This
temporal dependence disappears at either high interface velocities or large gap
spacings.

The power spectrum at saturation for five different velocities is shown in Fig.\
\ref{Fig:pow-sq-v}(b). The different behaviour at short and long length scales
is visible
in all cases, but the crossover from one regime to another (characterized by the
wavenumber $q_c$ at the crossing point of the two power laws) increases
systematically
with increasing velocities.

The exponent $\alpha_{1}$ (large $q$) increases with the interface velocity,
from a value
$\alpha_{1}= 0.9 \pm 0.1$ (slope $= -2.8 \pm 0.2$) at velocity $V$ to a value
$\alpha_{1}= 1.3 \pm 0.1$ (slope $= -3.6 \pm 0.2$) at velocity $10\,V$. We have
observed
that the variation is not linear, but tends to the limiting value
$\alpha_{1}=1.3$ at
large velocities.

The exponent $\alpha_{2}$ (small $q$) decreases with the interface velocity,
varying from
$\alpha_{2}= 0.6 \pm 0.1$ (slope $= -2.2 \pm 0.2$) at velocity $V$ to
$\alpha_{2}= 0.1
\pm 0.1$ (slope $= -1.2 \pm 0.2$) at velocity $10\,V$. This exponent seems to be
insensitive to the velocity at low velocities.

Fig.\ \ref{Fig:pow-sq-v}(c) shows the alternative analysis carried out to
determine the
roughness exponents, using $W(l)$. Notice that because this is a local analysis
it is
limited to a maximum value $\alpha=1$. At short length scales we obtain
$\alpha_{1}=0.90
\pm 0.1$ for all the velocities. The value $\alpha_{1}=1$ that we should have
obtained
for $v \geqslant 2V$ is unreachable because the fluctuations in $W(l)$
contribute always
in the direction of decreasing $\alpha_1$. At long length scales the values
obtained for
$\alpha_2$ are consistent with the values measured from the power spectrum.

\subsubsection{Effect of increasing the persistence of the disorder}

We have carried out a series of experiments oriented to explore how different
kinds of
disorder patterns affect the interfacial dynamics, and the possible universality
of our
results. The gap spacing and velocity used in all cases are $b=0.36$ mm and
$2V$.

\begin{figure*}
\includegraphics[width=17.8cm]{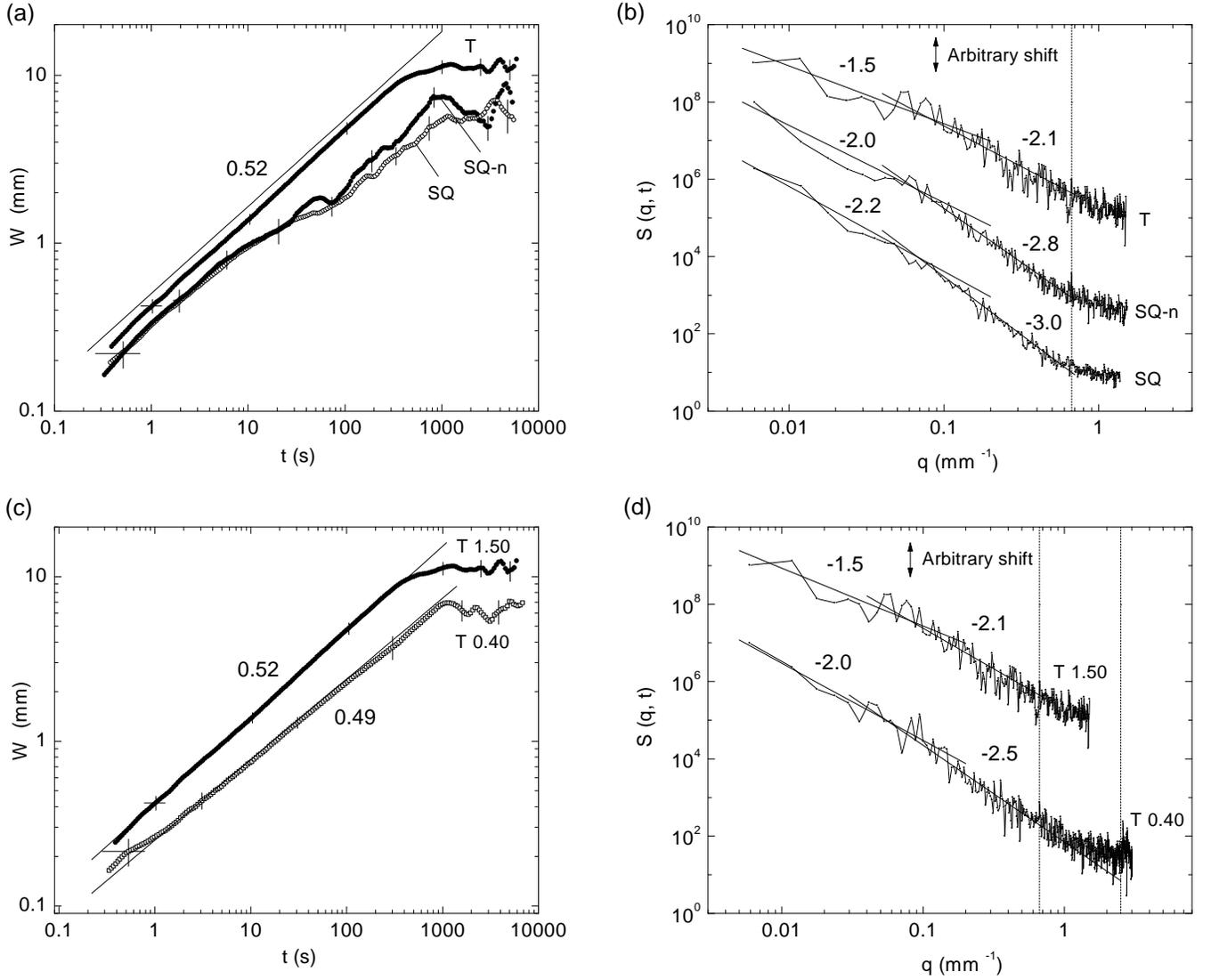}
\caption{Interfacial width $W(t)$ (a) and power spectra $S(q,t)$ (b) for three
kinds of
disorder with unit cell of size 1.50 mm. c) and d) show the same magnitudes for
disorder
T of two different sizes, 0.40 and 1.50 mm. The spectra have been shifted in the
vertical
direction for clearness. The vertical lines give the value of $q$ associated
with the
spatial scale of the disorder. In all the experiments $b=0.36$ mm and $v=2V$
(Ca'$= 3.25$
for T, and $3.69$ for SQ and SQ-n).} \label{Fig:noise-co}
\end{figure*}

When the correlation length of the disorder in the $y$ direction is increased
(changing
the disorder pattern from SQ to SQ-n and then to T), we observe important
differences in
the behaviour of both $W(t)$ and $S(q, t)$. The correlation length of the
disorder in the
$y$ direction is quantified through the average extent of the disorder cells in
the $y$
direction, $\widetilde{l}$, introduced in Sec.\ \ref{Subsec:disorder}. For a
basic
disorder cell of size 1.50 mm, $\widetilde{l} \simeq 2$ mm for SQ,
$\widetilde{l} \simeq
5$ mm for SQ-n, and $\widetilde{l}=\infty$ for T. The immediate consequence of a
progressively larger correlation length is the persistence of the capillary
forces for
longer times. The effect on $W(t)$ is that the large fluctuations observed
during growth
disappear as $\widetilde{l}$ increases. In the extreme limit of disorder T,
$W(t)$ is a
neat power law with an exponent $\beta=0.52 \pm 0.03$, as can be seen in Fig.\
\ref{Fig:noise-co}(a). Other relevant aspects are that the saturation times can
be
clearly identified and the fluctuations in saturation have a minor amplitude.
Whereas the
important fluctuations of $W(t)$ both during growth and saturation, together
with
substantial differences between different runs and disorder configurations, are
characteristic of the experiments with SQ, for T the experiments show that
different runs
with the same disorder configuration lead to practically identical results, and
differ
only slightly when the disorder configuration is changed.

The increasing correlation length in the direction of growth has important consequences
in the power spectrum. On one side, the observation of a smaller roughness exponent
$\alpha_{1}$ as time increases for disorder SQ (Fig.\ \ref{Fig:pow-sq-v}(a)), is more
evident as we go to SQ-n and T. The exponent $\alpha_1$ obtained for different disorders,
however, tends to a single value at large gap spacings and high velocities (weak
capillary forces). On the other side, the spectra at saturation shown in Fig.\
\ref{Fig:noise-co}(b), clearly indicate that a progressively larger $\widetilde{l}$
reduces the measured roughness exponents, from $\alpha_{1}=1.0 \pm 0.1$ to
$\alpha_{1}=0.5 \pm 0.1$, and from $\alpha_{2}=0.6 \pm 0.1$ to $\alpha_{2}=0.2 \pm 0.1$,
as the disorder changes from SQ to T. This trend is confirmed by experiments using SQ
0.40 ($\widetilde{l} \simeq 0.6$), where the evolution of $\alpha_1$ with time is not
observed due to the short $\widetilde{l}$ and, at saturation, gives exponents $\alpha_{1}
\simeq 1.1$ and $\alpha_{2} \simeq 0.5$.

Another interesting modification that can be introduced in the disorder is the
size of
the basic disorder cell. In addition to our usual experiments with lateral size
1.50 mm,
we have made experiments with disorder T of lateral size 0.40 mm. The $W(t)$
curves for
both cases are shown in Fig.\ \ref{Fig:noise-co}(c). Both curves give
approximately the
same growth exponent, around $\beta=0.50$. The small differences in slope around
this
value can be attributed to the reduced number of disorder configurations
studied. The
most noticeable difference is the saturation width, which is smaller for the
small
disorder size. The power spectrum at saturation for the two cases is shown in
Fig.\
\ref{Fig:noise-co}(d). Notice that in the case of T 0.40, the measured exponents
$\alpha_{1}$ and $\alpha_{2}$ increase considerably, and approach the values
measured for
SQ or SQ-n 1.50.

\subsubsection{T 1.50 mm}

Here we present a systematic study of the disorder T 1.50, in the same line as
for SQ
1.50 (Sec.\ \ref{Subsec:SQ}). The disorder T 1.50 ensures that the interface is
always
single--valued at any gap spacing and velocity, which allows exploring a wide
range of
experimental parameters.

\begin{figure}
\includegraphics[width=8.6cm]{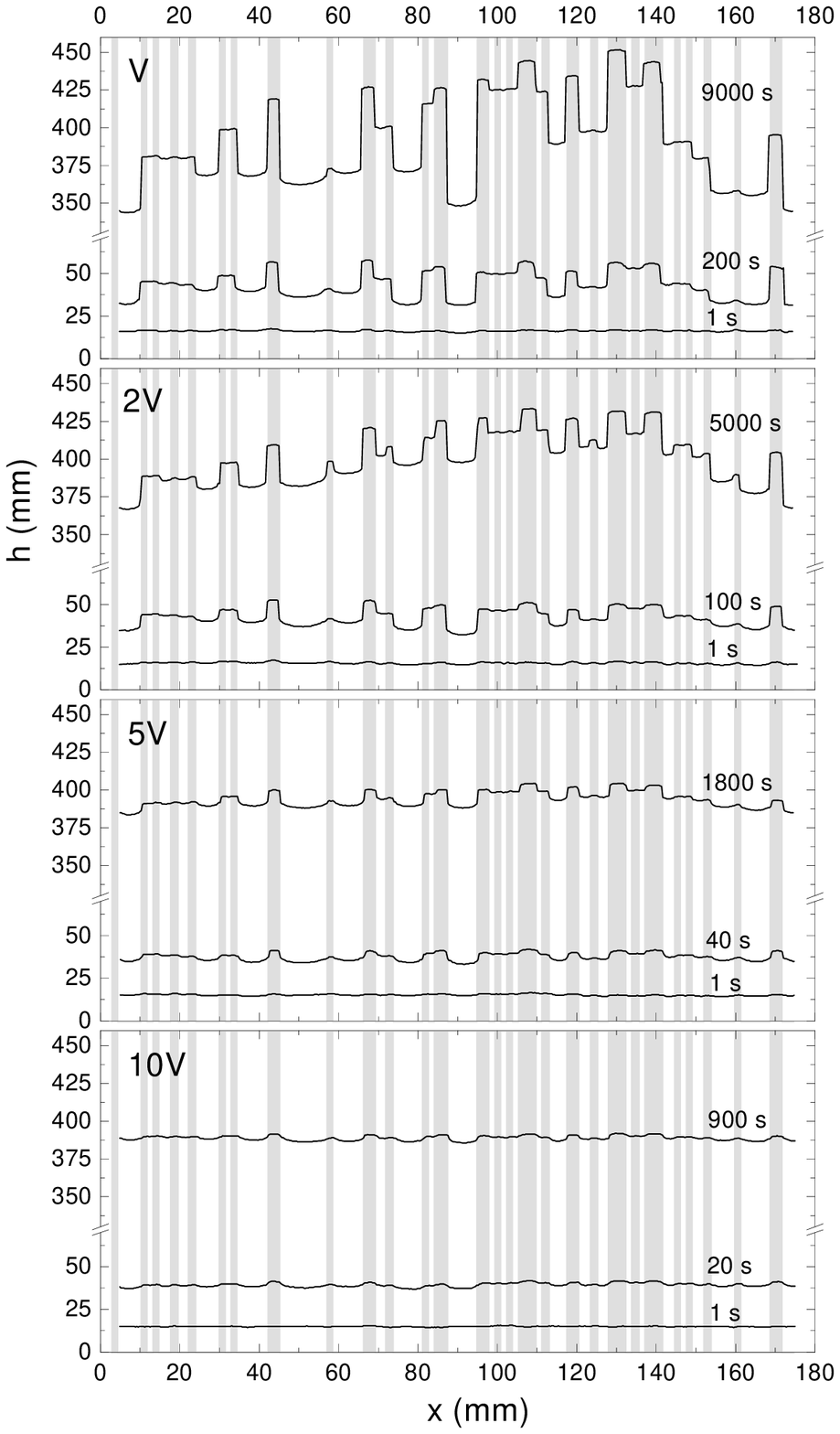}
\caption{Sequence of interfaces for disorder T 1.50 and gap spacing $b=0.36$ mm,
at
different velocities, corresponding to Ca'=$1.62$, $3.25$, $8.13$, and $16.3$.
The
disorder pattern is shown in grey.} \label{Fig:int-T-v}
\end{figure}

Fig.\ \ref{Fig:int-T-v} shows an example of the interfaces obtained in the
experiments
with T 1.50 at different velocities and fixed gap spacing. As we observed for SQ
1.50,
the interfaces become progressively flatter as the velocity is increased.
However, there
are interesting differences. The first one is the possibility to explore the
regime of
very low velocities, thanks to the fact that the continuity of the copper tracks
makes
local pinning impossible. However, there are two reasons for not selecting
velocities
lower than $V$. First, we have observed that for $v \lesssim V/4$ the oil
fingers become
so elongated that the interface pinches off at some distance behind the finger
tip. And,
second, because the oil tends to advance preferentially over the copper tracks,
at very
short times the oil over the fiber--glass recedes (due to mass conservation) and
parts of
the interface reach again the transverse copper tracks at the beginning of the
cell. This
limits the velocity to a minimum value $V$ for this gap spacing.

\begin{figure}
\includegraphics[width=8.6cm]{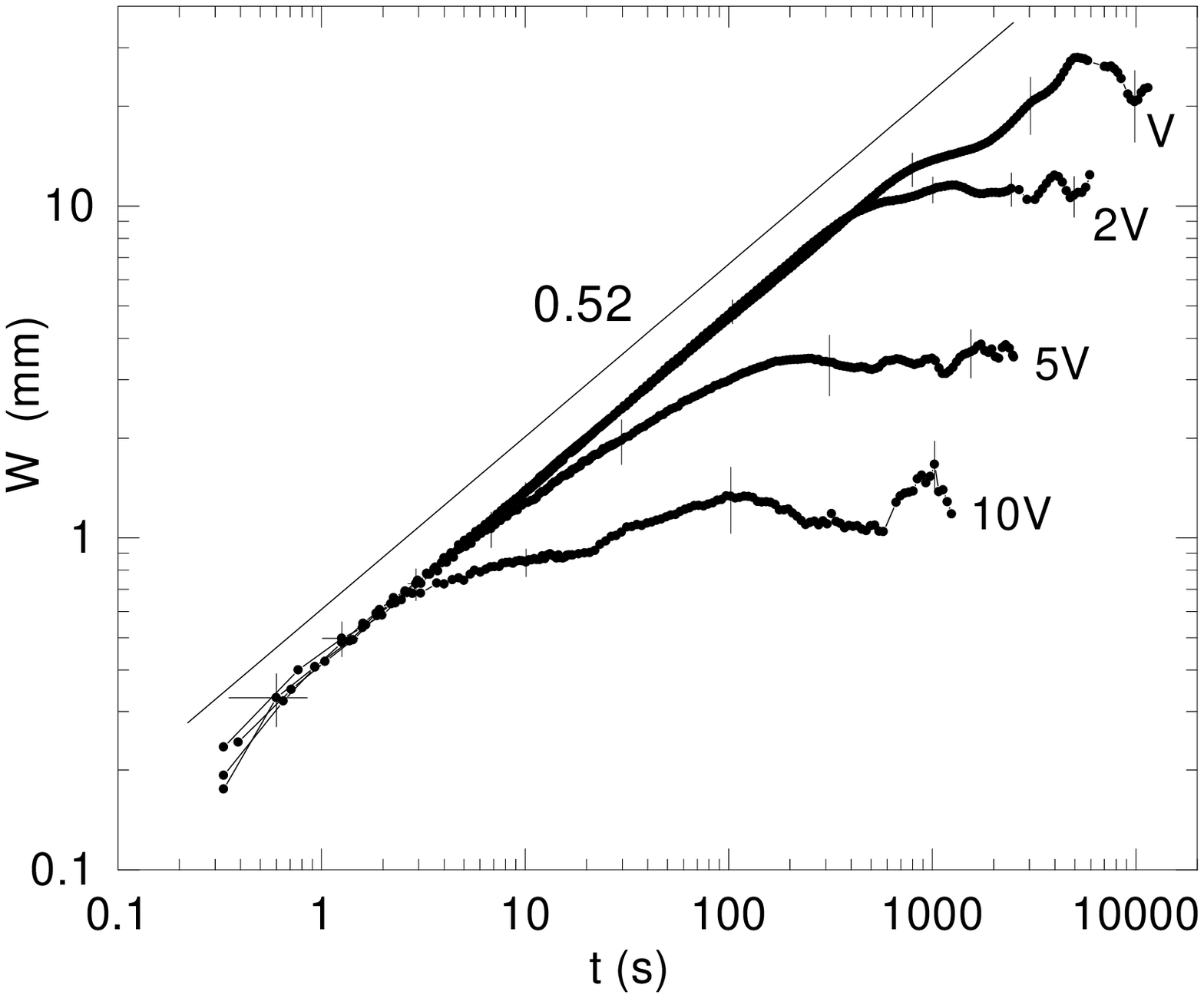}
\caption{Interfacial width,
$W(t)$, for disorder T 1.50, gap spacing $b=0.36$ mm, and four different
velocities. The
straight line drawn to guide the eye is the power law $W \sim
t^{0.52}$.}\label{Fig:t-velo}
\end{figure}

As discussed before, having tracks impedes that the interface gets locally
pinned. As a
result, the metastabilities observed for disorder SQ are not present here. Most
of the
fluctuations of $W(t)$ disappear, as shown in Fig.\ \ref{Fig:t-velo} for the
four
velocities studied. All the curves have approximately the same growth exponent
$\beta =
0.52 \pm 0.03$. The transition from the growth regime to the saturation regime
is rather
smooth and makes it difficult to measure the values of $t_{\times}$ and $W_{s}$.
Although
two identical runs (with the same disorder configuration) give almost identical
$W(t)$
curves, different disorder configurations give curves with slightly different
$\beta$,
$t_{\times}$, and $W_{s}$. In the former case the values vary between $\beta=
0.48$ and
$\beta= 0.55$. The fluctuations observed at saturation are attributed to small
inhomogeneities in the gap thickness, which affect particularly the experiments
at the
lowest velocities. Although the amplitude of the fluctuations decreases when we
increase
the number of disorder configurations, the almost perfect reproducibility of
experiments
with identical disorder configuration makes it difficult to have good statistics
using a
reduced number of disorder configurations.

\begin{figure}
\includegraphics[width=8.6cm]{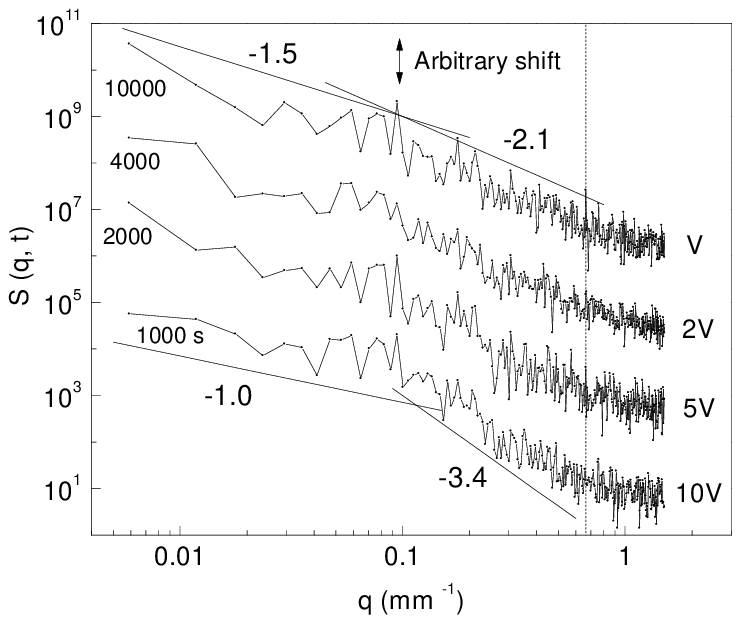}
\caption{Determination of the roughness exponents from the power spectra for
experiments
at four different velocities, with $b=0.36$ mm and disorder T 1.50. The curves
have been
shifted in the vertical direction for clearness. The vertical line gives the
value of $q$
associated with the spatial scale of the disorder. The other straight lines are
power law
fits from which the roughness exponents can be deduced.} \label{Fig:pow-t-v}
\end{figure}

The analysis of the power spectrum for the four velocities is shown in Fig.\
\ref{Fig:pow-t-v}. The deterministic character of the interfacial growth makes the
interfacial fluctuations not self--averaging. This leads to considerable fluctuations in
the power spectrum, that cannot be completely reduced without using a large number of
disorder configurations. Qualitatively, the power spectrum presents the same behaviour
observed for SQ, with two different regimes separated by a crossover point $q_c$. The
roughness exponent $\alpha_{1}$ increases with the velocity, varying from $\alpha_{1}=0.5
\pm 0.1$ (slope $-2.1 \pm 0.1$) for velocity $V$ to $\alpha_{1}=1.2 \pm 0.1$ (slope $-3.4
\pm 0.1$) for velocity $10\,V$. We have observed that at high velocities $\alpha_{1}$
reaches the same limiting value as it was measured for SQ 1.50.

\subsection{Variable gap spacing}

\subsubsection{SQ 1.50 mm}

\begin{figure}
\includegraphics[width=8.6cm]{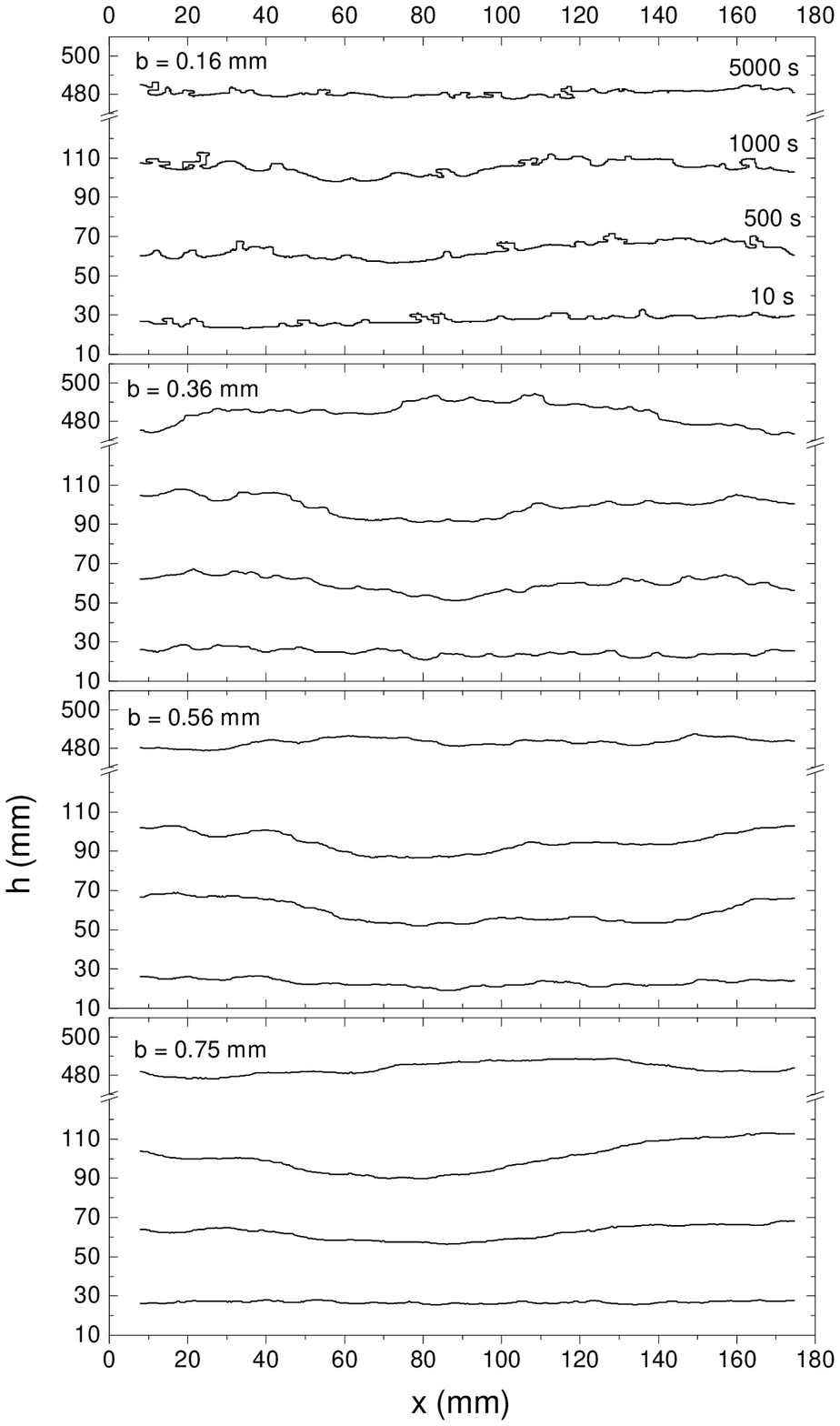}
\caption{Sequence of interfaces at velocity $2V$ and disorder SQ 1.50, at
different gap
spacings corresponding to, from smaller to larger gap spacing, Ca'$= 3.65$,
$3.70$,
$3.66$, and $3.62$. The disorder realization is the same in the four sequences.
The
reference velocity is $V=0.04$ mm/s.} \label{Fig:int-sq-g}
\end{figure}

Variations in the gap thickness modify the strength of the capillary forces in
the
vertical direction, which are destabilizing, in relation to the viscous forces
and to the
capillary forces in the plane of the cell, which are stabilizing. The sequence
of
interfaces at different $b$ and equal velocity ($2V$) of Fig.\
\ref{Fig:int-sq-g},
indicate that roughening at large scales is inhibited for the smallest gap
spacing,
$b=0.16$ mm, and the interface remains globally flat. Notice that the oil--air
interface
adapts perfectly to the disorder configuration. The gap spacings must be
increased to
values larger than $b=0.23$ mm to observe all scales contributing to the
roughening
process.

\begin{figure}
\includegraphics[width=8.6cm]{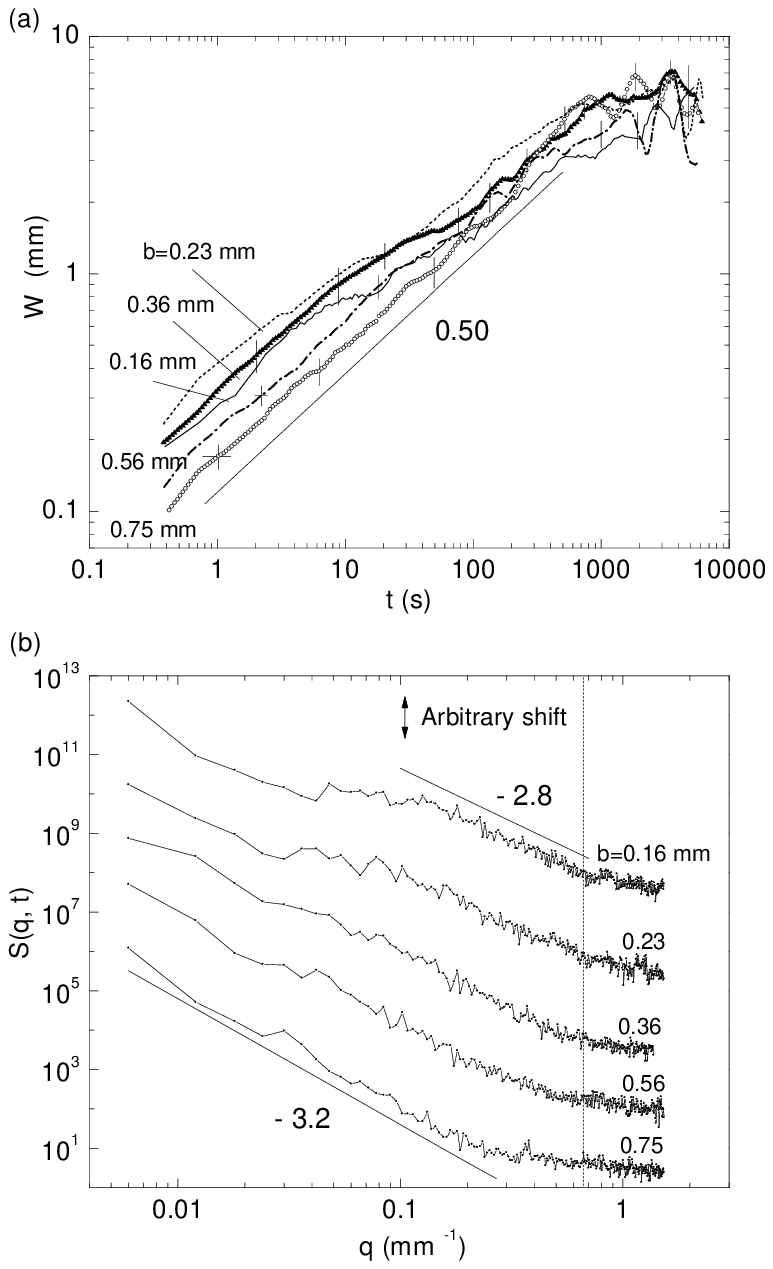}
\caption{Experimental results at velocity $2V$ and disorder SQ 1.50 at five
different gap
spacings. a) Interfacial width $W(t)$. A power law with slope $0.5$ has been
drawn to
guide the eye. b) Power spectra at saturation ($t=6000$ s). The curves have been
shifted
in the vertical direction for clearness. The vertical line gives the value of
$q$
associated with the spatial scale of the disorder.}\label{Fig:sq-gap}
\end{figure}

The results of $W(t)$ for different gap spacings are presented in Fig.\
\ref{Fig:sq-gap}(a). The results show that the saturation times and widths are
practically independent of gap spacing. This is a demonstration that increasing
the gap
thickness is not equivalent to increasing the velocity, although both changes
result in
larger Ca'. The growth exponent $\beta$ at the smallest gap spacings centers
around
$\beta=0.40$ but is subjected to a large error bar due to a significant presence
of
multivaluations in the interface. As $b$ increases there is a clear tendency
towards a
well defined power law, over three orders of magnitude in time, resulting in an
exponent
$\beta=0.50 \pm 0.02$.

Although all the experiments start with the same initial condition, the
transition from
an interface almost flat, with $W \simeq 0$, to a set of almost parallel $W(t)$
curves,
suggest that the growth at very short times is strongly dependent on gap
thickness.

The different role of velocity and gap spacing is also apparent from the power
spectrum
of the interfaces at saturation, shown in Fig.\ \ref{Fig:sq-gap}(b). For the
smallest gap
spacing, $b=0.16$ mm, the spectrum is nearly flat at small $q$, reflecting that
the
largest scales do not grow (Fig.\ \ref{Fig:int-sq-g}). The only contribution to
this part
of the spectrum comes from the memory of the global shape of the initial front,
which, at
the smallest gap spacings, cannot be prepared perfectly flat. The power law
dependence at
larger $q$ extends from $q_c \simeq 0.15$ to $q = 0.67$ (which corresponds to
the lateral
size of the disorder cell). As $b$ increases, the nearly $q$--independent
behaviour at
small $q$ changes to a power law behaviour (exponent $\alpha_1$), and the
crossover value
$q_c$ shifts to smaller $q$. For gap spacings as large as $b=0.75$ mm this shift
leads to
a single power law dependence, that could be strongly affected by finite--size
effects.
The measured exponents in the short length scale regime increase with $b$, from
$\alpha_{1}=0.9 \pm 0.1$ (slope $-2.8 \pm 0.2$) to $\alpha_{1}=1.1 \pm 0.1$
(slope $-3.2
\pm 0.2$).

\subsubsection{T 1.50 mm}

\begin{figure}
\includegraphics[width=8.6cm]{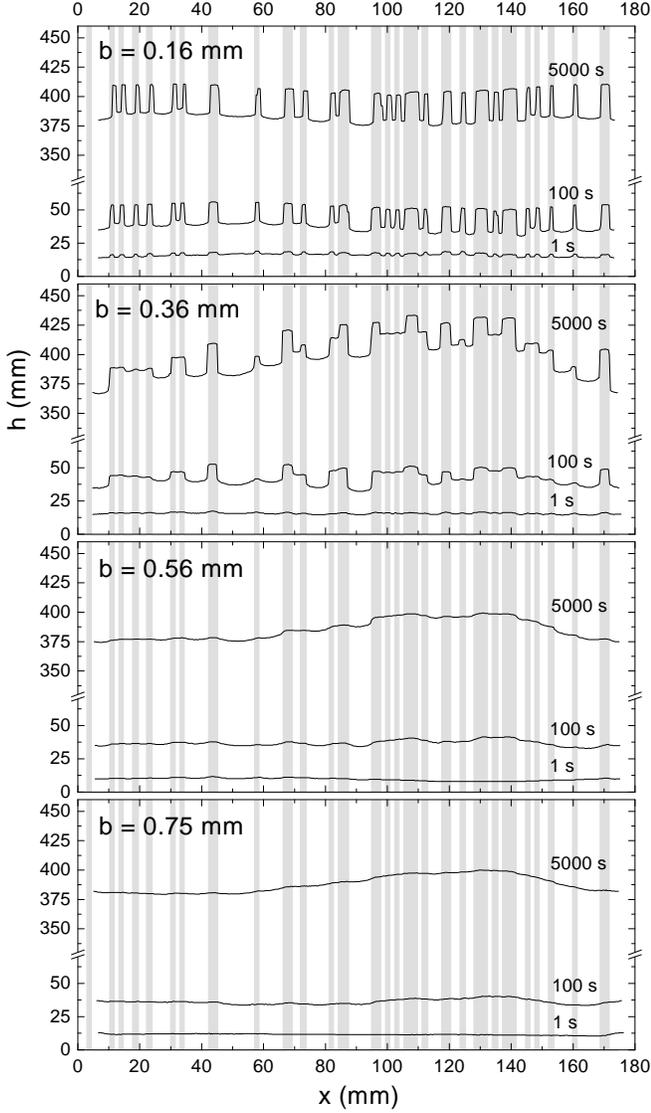}
\caption{Sequence of interfaces at velocity $2V$ and disorder T 1.50, at
different gap
spacings corresponding to, from smaller to larger gap spacing, Ca'$= 2.67$,
$3.25$,
$3.38$, and $3.41$.} \label{Fig:int-t-g}
\end{figure}

The gap spacing has been varied in the range $0.16 \leqslant b \leqslant 0.75$,
keeping
the velocity fixed at $2V$. Fig.\ \ref{Fig:int-t-g} show a sequence of the
temporal
evolution of the interfaces for four different gap spacings. The most remarkable
aspect
is the significant difference in the shape of the interfaces between $b=0.16$ mm
and the
other gap spacings. For $b=0.16$ mm the capillary forces are strong enough to
inhibit
correlations between neighbouring tracks. Similarly to the experiments with SQ
1.50 and
the smallest gap, we do not expect saturation of scales larger than the average
lateral
size of the tracks. For $b \geqslant 0.36$ mm the behaviour is different: the
interfacial
height is correlated from the size of the basic disorder cell up to the system
width. The
interface at long length scales adopts a final shape at saturation identical for
the
three gaps, dominated by the spatial distribution of the disorder pattern. At
short
length scales the amplitude of the fingers depends on the relative strength of
the
capillary forces, tuned by the gap spacing.

\begin{figure}
\includegraphics[width=8.6cm]{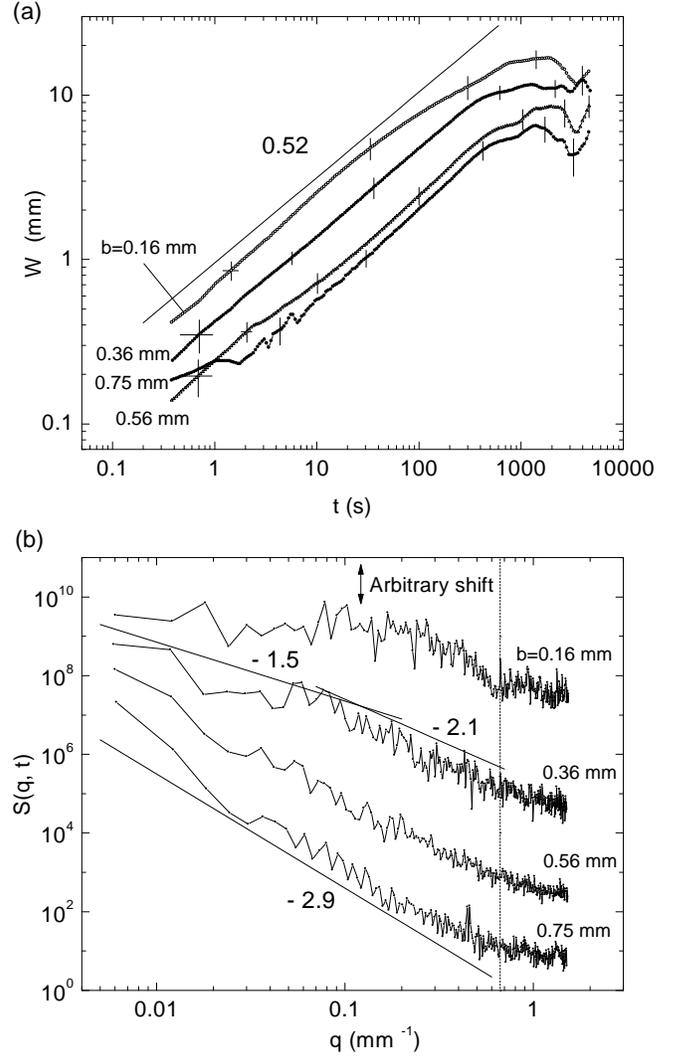}
\caption{Experimental results at velocity $2V$ and disorder T 1.50, at four
different gap spacings. a) Interfacial
width $W(t)$. A power law with slope $0.52$ has been drawn to guide the eye. b)
Power
spectra at saturation ($t=6000$ s). The curves have been shifted in the vertical
direction for clearness. The vertical line gives the value of $q$ associated
with the
spatial scale of the disorder.} \label{Fig:t-gap}
\end{figure}

Fig.\ \ref{Fig:t-gap}(a) shows the $W(t)$ curves for the four gap spacings. The
slopes
are practically the same in all four cases, the saturation times are also very
similar,
and the saturation widths are progressively larger as the gap spacing is
reduced. The
fact that the interfacial fluctuations with $b=0.16$ mm reach saturation is a
consequence
of driving the system with a finite velocity that impedes an infinite
deformation of the
interface.

In Fig.\ \ref{Fig:t-gap}(b) we represent the spectra for the four velocities at
saturation. The fact that for $b=0.16$ mm the interfacial height is uncorrelated for
distances larger than the average lateral size of the tracks ($2.3$ mm, corresponding to
$q=0.43$) is clearly observed, since for $q \lesssim 0.4$ we observe a {\it plateau} in
the power spectrum. The only hint of a power law regime is at large $q$, from $q \gtrsim
0.4$ to the limit given by the disorder size, an interval too short to measure any
roughness exponent. The other gap spacings show qualitatively the same behaviour observed
for SQ. The large $q$ saturate with an exponent $\alpha_{1}$ that is progressively larger
as we increase the gap spacing, varying from $\alpha_{1}=0.5 \pm 0.1$ (slope $-2.1 \pm
0.2$) for $b=0.36$ mm to $\alpha_{1}=0.9 \pm 0.1$ (slope $-2.9 \pm 0.2$) for $b=0.75$ mm.
The regime characterized by $\alpha_{2}$ is only identifiable for $b=0.36$ mm. For the
other gap spacings, our results indicate that we have a unique power law over all length
scales. Here again, however, the results for the largest gap spacings could be severely
influenced by finite size effects.

\section{Analysis and discussion}\label{Sec:Analysis}

\begin{figure}
\includegraphics[width=8.6cm]{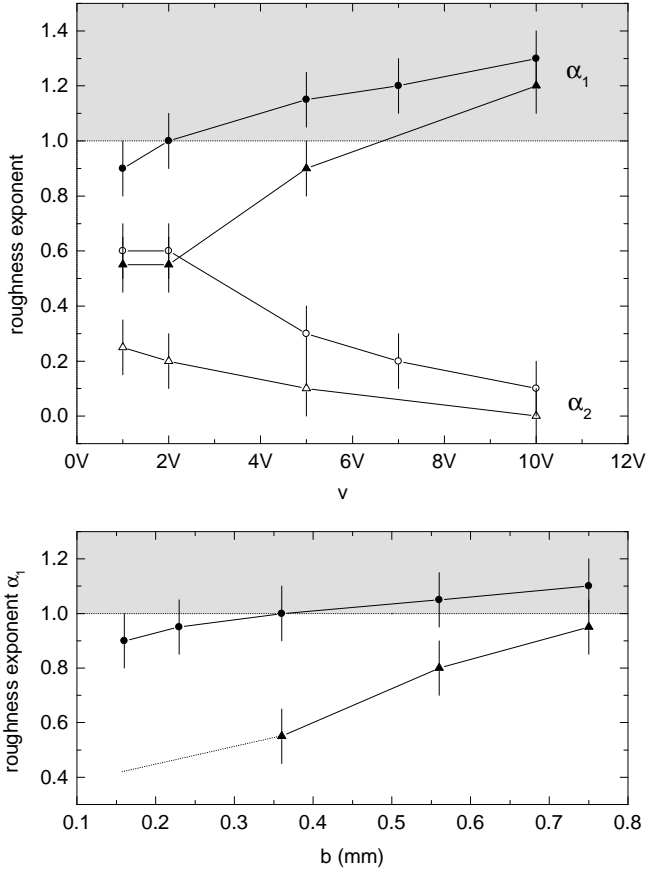}
\caption{Variation of the roughness exponents with interface velocity (top) and gap
spacing (bottom), for SQ 1.50 (circles) and T 1.50 (triangles). Solid and open symbols
correspond to $\alpha_{1}$ and $\alpha_{2}$, respectively. The region of super--rough
interfaces ($\alpha > 1$) is shown in grey. The lines are drawn to guide the eye.}
\label{Fig:alfa-vel}
\end{figure}

The variation of the roughness exponents with velocity and gap spacing, for SQ
and T
disorder configurations, is summarized in Fig.\ \ref{Fig:alfa-vel}. We observe
that
$\alpha_1$ and $\alpha_2$ have opposite behaviours as the interface velocity
increases.

The exponent $\alpha_1$, characteristic of roughening at short length scales, increases
with $v$. It becomes larger than $1$ (super--rough interface) already at moderate
velocities. To understand this behaviour, we refer to Fig.\ \ref{Fig:pow-sq-v}(a), where
it is shown that $\alpha_1$ is large at very short times, and decreases progressively as
time goes on, at velocity $2V$. The initial super--roughness is due to the initially
local dynamics of the interface fluctuations as the interface gets in contact with the
disorder for the first time. At sufficiently large velocities, however, the subsequent
decrease of $\alpha_1$ is not observed, and the initial super--roughness gets frozen in.
The reason is that the saturation time $t_{\times}$ is comparable to the average time
spent by the interface to go across the distance $\widetilde{l}$ (the average length of
the disorder in the $y$ direction), estimated as $\tau = \widetilde{l}/v$. For example,
for SQ 1.50, $b=0.36$ mm and $v=10\,V$, we have $t_{\times}=5$ s and
$\tau=\widetilde{l}/v \simeq 6$ s. This explains also that $\alpha_{1}$ takes similar
values for SQ and T at high velocities, since whenever $t_{\times} \simeq \tau$ the
continuity of the disorder in the $y$ direction is irrelevant. In particular, we have
evidence that $\alpha_1$ becomes identical for SQ and T at very large velocities. The
interface is saturated at all time, making the interfacial dynamics in SQ equivalent to
an average of the dynamics over a number of T configurations. For gap spacings around
$b=0.36$ mm, we have observed that this condition is satisfied for $v \gtrsim 13V$. An
identical value $\alpha_{1} \simeq 1.3$ has been obtained for both SQ and T by performing
experiments at $30V$ ($t_{\times} \ll \tau$). This value of $\alpha_1$ is distinctively
larger than previous experimental results, and coincides within error bars with the value
$1.25$ found numerically in  Ref.~\cite{Dube-teoric-00} taking account of the
nonlinearities. A super--rough interface ($\alpha > 1$) is also obtained by scaling
analysis in \cite{Aurora-EPL-01}, but with a larger exponent.

\begin{figure}
\includegraphics[width=8.6cm]{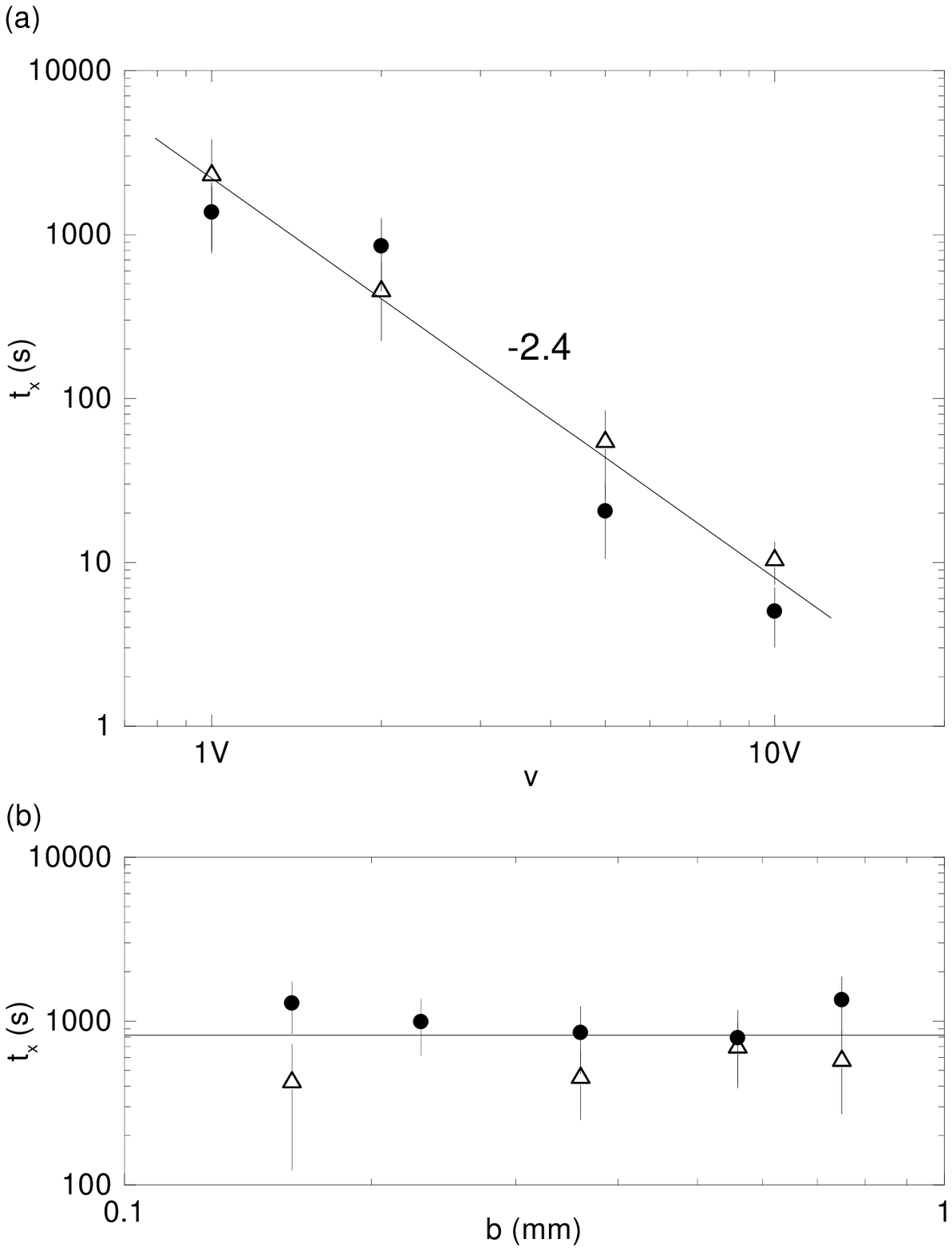}
\caption{Saturation time $t_{\times}$ as a function of velocity
$v$ (top) and gap spacing $b$ (bottom), for SQ 1.50 (circles) and T 1.50
(triangles). The
straight lines have been drawn to guide the eye.}\label{Fig:tx-v}
\end{figure}

\begin{figure}
\includegraphics[width=8.6cm]{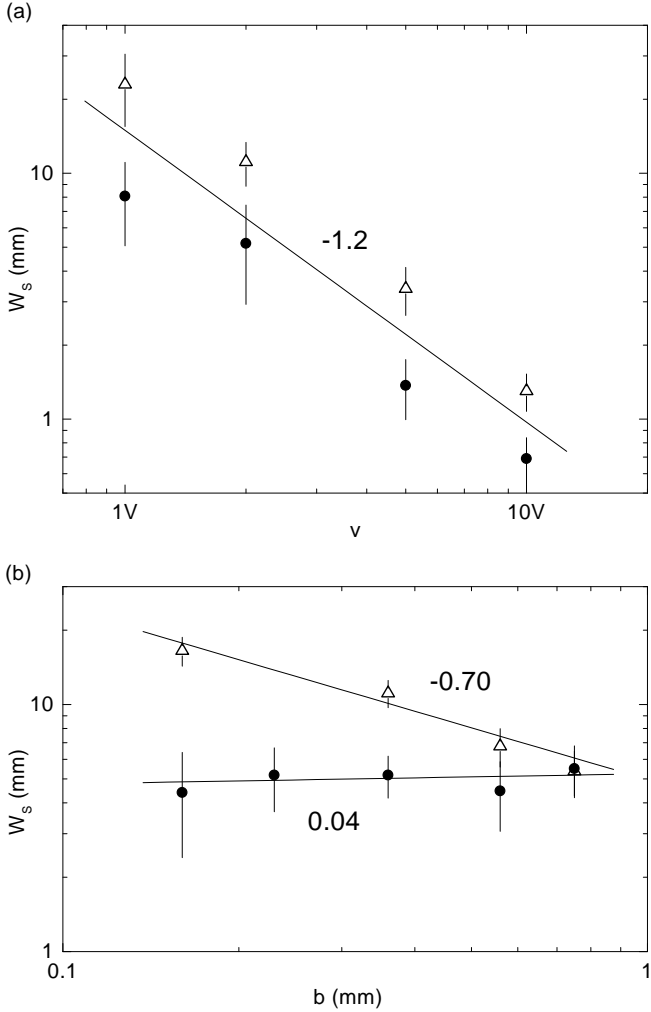}
\caption{Saturation width $W_s$ as a function of velocity $v$ (top) and gap
spacing $b$
(bottom), for SQ 1.50 (circles) and T 1.50 (triangles). The straight lines have
been
drawn to guide the eye.} \label{Fig:ws-v}
\end{figure}

\begin{figure}
\includegraphics[width=8.6cm]{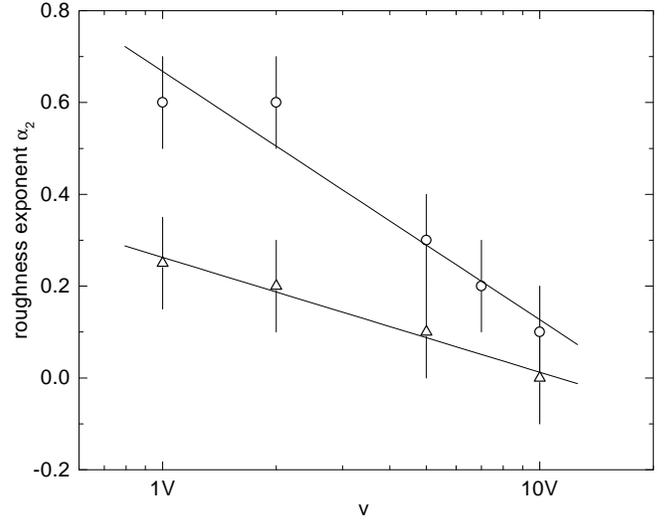}
\caption{Roughness exponent $\alpha_{2}$ as a function of $\log v$ for SQ 1.50
(circles)
and T 1.50 (triangles). The straight lines have been drawn to guide the eye.}
\label{Fig:alfa-logv}
\end{figure}

The exponent $\alpha_{2}$ (long length scales) measured in our experiments
decreases with
$v$. Its actual behaviour, however, could be obscured by finite size effects.
The reason
is that nonlocal effects on the interfacial dynamics (due to the fluid flow from
the
reservoir) become progressively more significant as $v$ increases. For
experimental
reasons we have not studied finite size effects in any systematic way. However,
we do
have checked that experiments performed at velocity $V$ and system size $L/2$
give the
same exponent $\alpha_2$ than those at velocity $2V$ and system size $L$. The
simultaneous dependence of $\alpha_{2}$ on velocity and system size can be
estimated from
$W(t)$ for both SQ and T. We have observed (Figs.\ \ref{Fig:tx-v}(a) and
\ref{Fig:ws-v}(a)) that the saturation time $t_{\times}$ and the saturation
width $W_{s}$
are power laws of the velocity $v$, of the form $t_{\times} \sim v^{-\delta}$
and $W_{s}
\sim v^{-\gamma}$, with $\delta\simeq 2.5$ and $\gamma \simeq 1.1$ for SQ, and
$\delta\simeq 2.3$ and $\gamma \simeq 1.2$ for T. Combining these relations with
the FV
scaling assumption (\ref{Eq:FV}), and taking into account that the saturation
times and
widths correspond to the long time regime (long length scales), we get:
\begin{equation}\label{alfa-v}
  \alpha_{2} \sim -\beta \delta \frac{\log v}{\log L},
\end{equation}
with $\gamma=\beta\delta$. Given that the system size $L$ has been kept fixed in
the
experiments, the relation (\ref{alfa-v}) predicts that the roughness exponent
$\alpha_2$
obtained from the experiments will depend on $v$. The validity of this result is
confirmed in Fig.\ \ref{Fig:alfa-logv}, where the values of $\alpha_2$ are
plotted vs.
$\log v$.

From the observation that the difference between $\alpha_1$ and $\alpha_2$ decreases as
$v$ is reduced (Fig.\ \ref{Fig:alfa-vel}), and taking into account that capillary forces
become also more important, we argue that in the limit of very low velocities the power
spectrum would display an unique power law extending over all scales with a roughness
exponent in the interval $0.6-0.9$, compatible with the observations in
\cite{he-92,Rubio-89,Horvath-90,Rubio-90,Horvath-91}. This limit is unreachable for SQ
because at velocities $v \lesssim V/2$ the interface gets locally pinned and develops
overhangs. In the case of T there is no pinning or multivaluation, but the interface
stretches to such a point that easily pinches off at long times, before saturation. These
and other questions, present only in the regime of flows dominated by capillary forces,
are studied in Ref.\ \cite{Soriano-2001-II}.

Concerning the influence of $b$, the gap spacing, two main effects must be taken
into
account. First, the mobility is proportional to $b^2$ and hence increases with
$b$,
enhancing the stabilizing effect of the viscous pressure field (on long length
scales)
and the interfacial tension in the plane of the cell (on short length scales).
Second,
increasing $b$ weakens the destabilizing effect of the capillary forces induced
by the
disorder, which act on short length scales. The overall result is that the
curves $W(t)$
are shifted towards smaller widths as $b$ increases, as shown in Figs.\
\ref{Fig:sq-gap}(a) and \ref{Fig:t-gap}(a). The shift is such that $W\cdot b$
collapses
the curves for different gap spacings into a single one, in both SQ and T cases.
The fact
that large values of $b$ makes the presence of the disorder irrelevant can be
observed
through the values of the roughness exponents $\alpha$, which show a tendency to
converge
into a single value $\alpha \simeq 1$ at large gap spacings and for any disorder
configuration.

Another observation is that the saturation regime is reached always at almost
the same
saturation time $t_{\times}$, independently of the gap spacing, as Fig.\
\ref{Fig:tx-v}(b) reveals. The interfacial width at saturation, however, has a
different
origin depending on gap spacing. For small $b$ the width is due to fluctuations
at the
shortest length scales, while for large $b$ it is due to fluctuations at longer
length
scales. This can be observed on the interfaces and is clearly reflected on the
power
spectrum (Figs.\ \ref{Fig:sq-gap}(b) and \ref{Fig:t-gap}(b)). The passage from
short to
long length scales is continuous with $b$. It can be shown that replacing the
variable
$q$ by the dimensionless variable $q \cdot b$, produces the collapse of the
power law
region of the power spectra in Fig.\ \ref{Fig:sq-gap}(b).

\begin{table*}
\begin{center}
\begin{tabular}{|c|cc|cc|cc|}
\hline Disorder & \multicolumn{2}{c|}{Low Ca'} & \multicolumn{2}{c|}{Moderate
Ca'} &
\multicolumn{2}{c|}{Large Ca'}
\\ & \multicolumn{2}{c|}{\;($\lesssim  3$)\;}
& \multicolumn{2}{c|}{\;($3 < $ Ca' $< 10$\;} & \multicolumn{2}{c|}{\;($\gtrsim
10$)\;}

\\ \hline \hline & $\alpha_1$ & $\alpha_2$ & $\alpha_1$ & $\alpha_2$ &
$\alpha_1$ & $\alpha_2$
\\ \hline \;SQ 1.50\; & $0.6-0.9$ & $\--$ & $\simeq 1$ & $\simeq 0.6$
& $\simeq 1.3$ & $\simeq 0$
\\ \hline \;SQ-n 1.50\; & $0.6-0.9$ & $\--$ & $\simeq 0.9$ & $\simeq 0.5$
& $\simeq 1.3$ & $\simeq 0$
\\ \hline \;T 0.40\; & \multicolumn{2}{c|}{New regime} & $\simeq 0.7$ & $\simeq
0.5$
& $\simeq 1.3$ & $\simeq 0$
\\ \hline \;T 1.50\; & \multicolumn{2}{c|}{New regime}  & $\simeq 0.5$ & $\simeq
0.2$
& $\simeq 1.3$ & $\simeq 0$
\\ \hline
\end{tabular}
\caption{Summary of the values of the roughness exponents.}
\label{TAB:main-results}
\end{center}
\end{table*}

Table\ \ref{TAB:main-results} shows the main results obtained for the roughness exponents
$\alpha_1$ (short length scales) and $\alpha_2$ (long length scales) at different
disorder configurations and drivings. For small Ca', capillary forces are dominant at all
scales, and the dynamics is very sensitive to the disorder configuration. For SQ and
SQ-n, the interfaces get locally pinned, and the measured exponents are close to those
obtained in DPD or in experiments where capillary forces are dominant
\cite{he-92,Rubio-89,Horvath-90,Rubio-90,Horvath-91}. In the limit of persistent
disorder, the nature of the disorder impedes pinning, but the effect of the destabilizing
capillary forces combined with the correlations between neighboring tracks leads to a new
regime that can be described using the anomalous scaling ansatz \cite{Soriano-2001-II}.
This new regime also extends to the region of moderate Ca' for T disorder. For moderate
Ca', viscous forces are dominant at long length scales and two clear regimes separated by
a crossover wavenumber $q_c$ can be characterized. For SQ and SQ-n we get roughness
exponents $\alpha_1 \gtrsim 1$ and $\alpha_2 \simeq 0.5 - 0.6$. For T, we get a
qualitatively similar behaviour, but obtaining lower values of the exponents due to
finite size effects at long length scales and the dominant capillary forces at short
length scales. For large Ca', the viscous forces cause that the initial super--roughness
at short times and short length scales gets frozen in, obtaining the same roughness
exponent $\alpha_1 \simeq 1.3$ for all the disorder configurations.

Our experimental results can be compared now with the predictions of the nonlocal models
\cite{ganesan-98, Dube-teoric-99,Dube-teoric-00,Aurora-EPL-01} discussed in Section II.
It appears that none of the models can account for the different results obtained in the
whole range of capillary numbers explored.  Although the Flory--type argument used by
Ganesan and Brenner \cite{ganesan-98} is difficult to justify in this nonequilibrium
situation, they obtain roughness exponents ($\alpha_1 = 3/4$ and $\alpha_2 = 1/2$)
consistent with our values for small and moderate Ca'.  The exponents obtained by
Hern\'andez--Machado {\it et al.} \cite{Aurora-EPL-01} for the long length scales at
moderate Ca', $\alpha_2=1/2$ and $z=1$ ($\beta_2=1/2$), are in agreement with our
experimental values in the long time regime, where the viscous forces are dominant.  The
value of $\alpha$ for short length scales, corresponding to the short time regime of the
model, gives a super--rough behaviour ($\alpha_1 > 1$) dominated by surface tension in
the plane, but larger than the exponent measured in the experiment.  For large capillary
numbers the numerical result of Dub{\'e} {\it et al.}
\cite{Dube-teoric-99,Dube-teoric-00}, $\alpha_1 = 1.3$, is in agreement with our
experimental results, but not the result $\beta_1=0.3$ because, as pointed out above,
these authors assume a spontaneous imbibition.  At this point it would be interesting to
know whether a model containing a quenched disorder both in the mobility and in the
chemical potential would be able to explain the experimental results for the whole regime
of Ca' studied here, and specially the regime of large Ca' in forced imbibition. This
remains an open question.

Our last analysis is the variation of $q_c$ with $v$. The crossover wavenumber $q_c$
separates the regime in which the long length scale fluctuations (small $q$) are damped
by the viscous pressure field, from the regime in which the short length scale
fluctuations (large $q$) are damped by the interfacial tension in the plane of the cell.
Since the relative importance of the viscous pressure field increases with Ca', it is
expected that $q_c$ will increase with $v$. Specifically, a linear analysis of the
interfacial problem shows that the viscous damping is proportional to $v |q|$ and the
interfacial tension damping is proportional to $q^2 |q|$, which results in $q_c \propto
v^{0.5}$ \cite{Dube-teoric-00,Aurora-EPL-01}.

\begin{figure}
\includegraphics[width=8.6cm]{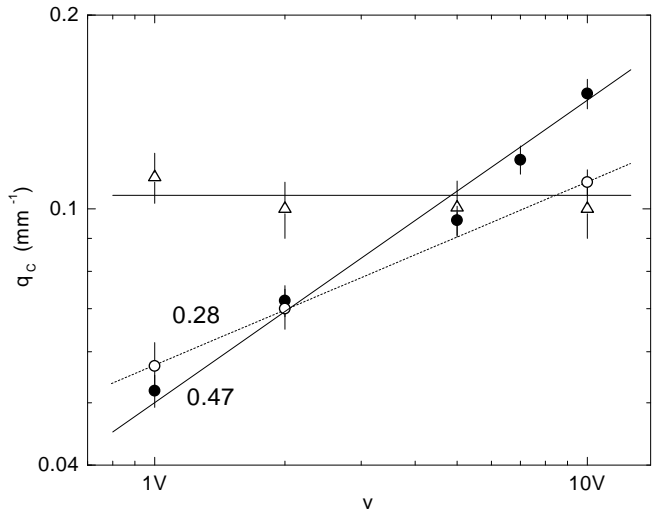}
\caption{Dependence of the crossover
wavenumber $q_{c}$ on velocity $v$ for three disorder configurations: SQ 1.50
(solid
circles), SQ-n 1.50 (open circles) and T 1.50 (triangles).} \label{Fig:qx}
\end{figure}

In Fig.\ \ref{Fig:qx} we show the behaviour of $q_c$ with $v$ for different
kinds of
disorder. The error bars are relatively large because the exact location of
$q_c$ in the
experimental power spectra is difficult to ascertain. For SQ 1.50 we find that
$q_c
\propto v^{0.47}$, in good agreement with the theoretical prediction. It is
interesting
to note that, as we go to disorders of larger $\widetilde{l}$ (increasing
persistence),
$q_{c}$ tends to be less sensitive to $v$. For T 1.50 we observe that $q_c$
becomes
independent of $v$, within error bars. The reason can be understood in the
framework of
the following scenario: at a local level (see Fig.\ \ref{Fig:close-up}) the
motion of the
interface in the SQ disorder can be viewed as a series of events formed by a
period of
nearly steady motion, a subsequent period of fast advance over a copper island
(accompanied by an abrupt change of sign of the in--plane curvature), and a
third period
of fast advance over fiber--glass due to relaxation of the local in--plane
curvature.
Thus, damping of the short length scales due to the interfacial tension in the
plane of
the cell is basically effective only in this last period, when the interface
depins from
the copper islands. As the persistence of the disorder in the $y$ direction is
larger,
the relaxation periods are less frequent. This can be observed through the
histogram of
Fig.\ \ref{Fig:noise}. When we increase the persistence of the disorder changing
from SQ
to SQ-n, the number of the smallest copper aggregations reduces almost one order
of
magnitude. In the limiting case of T 1.50 the disorder is continuous in the $y$
direction, and the damping role of the in--plane interfacial tension is
effectively
suppressed.

\section{Conclusions}\label{Sec:Conclusions}

We have presented new experiments of forced fluid imbibition in a Hele--Shaw
cell with
quenched disorder. We have used three main kinds of disorder patterns, SQ, SQ-n
and T,
characterized by an increasing persistence length $\widetilde{l}$ in the
direction of
growth. We have measured a robust roughness exponent $\beta \simeq 0.5$ that is
almost
independent of the disorder configuration, interface velocity, and gap spacing,
although
the behaviour of the interfacial width presents important fluctuations both
during growth
and at saturation, that progressively disappear as the disorder is more
persistent in the
$y$ direction. The roughness exponent $\alpha$, however, shows a clear
dependence on the
experimental parameters, as summarized in Table\ \ref{TAB:main-results},
discussed in the
previous Section. Finally, we have focused on the dependence of the crossover
wavenumber
as a function of the interface velocity $v$ and $\widetilde{l}$. For the
shortest
$\widetilde{l}$, $q_c \sim v^{0.47}$, and becomes independent of $v$ as the
disorder is
persistent in the $y$ direction.

The absence of pinning in experiments with disorders of increasing $\widetilde{l}$ allows
a detailed investigation of the regime of large capillary forces. This regime presents
some interesting novel features which are studied in detail in Ref.\
\cite{Soriano-2001-II}.


\section{Acknowledgements}
We are grateful to M. A. Rodr\'{\i}guez, L. Ram\'{\i}rez-Piscina, J. Casademunt,
K. J.
M{\aa}l{\o}y, and J. Schmittbuhl for fruitful discussions. The research has
received
financial support from the Direcci\'on General de Investigaci\'on (MCT, Spain),
project
BFM2000-0628-C03-01. J. O. acknowledges the Generalitat de Catalunya for
additional
financial support. J. S. is supported also by a fellowship of the DGI (MCT,
Spain).


\end{document}